# High-performance neuromorphic computing architecture of brain

Jinxuan Ma[1], Wanlin Guo[1,2]*

**Affiliations:**

[1]State Key Laboratory of Mechanics and Control for Aerospace Structures, Nanjing University of Aeronautics and Astronautics, Nanjing, Jiangsu, China.

[2]Institute for Frontier Science of Nanjing University of Aeronautics and Astronautics, Nanjing University of Aeronautics and Astronautics, Nanjing, Jiangsu, China.

*Corresponding author. Email: wlguo@nuaa.edu.cn

**Abstract:** Artificial intelligence can outperform humans in specific tasks but consumes substantial energy. How the human brain can work at just 20 watts with complex cognitive intelligence? Here we decode the fundamental information strategy unit of brain, neural sphere, which agglomerates neurons into sphere to achieve energy-efficient and exhibits many ultra-long period or random electrophysiological activities. Chaos dynamics and fractal theory demonstrated the mathematical principle of neural spheres to memorize and process through different electrophysiological activities which depend on strange attractors. A high-performance neuromorphic computing architecture of brain was then constructed which predicts a storage capacity of $7.48\times10^{18}$ Bytes and a computational power of $6.24\times10^{18}$ FLOPS for human brain. At this capacity, the energy efficiency of the human brain after long-term evolution can be up to 79% via Landauer's principle, 8-order higher than that of the latest computer chips, supporting the rationality of the proposed architecture.

## Introduction

People have long been committed to the development of brain science to inspire the design of Artificial General Intelligence serving production and life. In 2016, the groundbreaking emergence of AlphaGo(*1*) sparked an artificial intelligence (AI) storm that continues to this day. Relying on trillions of nanoscale field-effect transistors, AI systems process data and compute at a frequency of several gigahertz, but at the cost of huge energy consumption. In comparison, the human brain consists of 86 billion micron scale neurons that transmit and process information through action potential spikes on the order of milliseconds. It achieves remarkable cognitive feats with a mere 20 watts of metabolic power(*2*). Understanding the mysterious and delicate working mechanisms of the brain to develop brain-inspired computing technologies has remained an extraordinary challenge worthy of long-term exploration(*3, 4*).

Pioneering neuroscience research has dissected the signal transmission function of neurons and the connection structure of the brain. On the basis of Cajal's elucidation of the independence of neurons in 1888(*5*), Hodgkin and Huxley summarized the ionic mechanisms underlying the generation and propagation of action potentials in 1950s(*6-10*). This breakthrough model and its variants(*11, 12*) construct the electrophysiological mechanisms of microscopic neural dynamics. With the development in technology of electron microscopy, magnetic resonance imaging(*13*), multi-channel electrodes(*14*), two-photon imaging(*15*) and other technologies, a large number of neural structures and electrophysiological activities(*16*) from ion channels, synapses, neural networks to cortex have been recorded, providing an increasing data support for the unification of microscopic electrical signal transmission and macroscopic neural network calculation principle. Large amounts of neuronal morphology and neural connection data have been cataloged in databases such as Digitally Reconstructed Neurons and Glia(*17*). Some works had characterized the neural network properties of connectomes in various species, including the nervous system of Caenorhabditis elegans(*18*), the whole-brain map of drosophila(*19*), partial cortical maps of mice(*20, 21*), the mesoscale connectome of the human brain(*22*), etc. The Human Brain Project(*23*) in Europe has also made great efforts in this regard. Based on the analysis of the network structure of the connectome, repeated neuronal connectivity patterns in different brain regions at different spatial scales have been found, including neuronal motifs(*24*), lager-scale architectural plan(*25*), etc. Moreover, existing studies have estimated that the theoretical storage capacity of the human brain may be between 500 TB and 1 PB based on neural connect-omics and synaptic polymorphism(*26*). However, it remains a great challenge to understand how the neural system store and process information from the huge information flow, heterogeneous networks and highly nonlinear dynamics of connect-omics. The existing study mainly focuses on the biological phenomena, but the operation mechanism of the brain remains elusive.

Extracting information patterns from the reconstructed electrophysiological activity is an effective way to explore the possible working mechanism of the brain. Here we developed a neural sphere by optimizing the neural distribution with the objective of minimum energy consumption and parameterizing the neural microstructures and electrophysiological functions to simulate the action potential activities of the brain. Based on the comprehensive simulation, it is found that specially designed small-scale neural spheres have the characteristics of ultra-long period action potential activity exceeding 100 ms or aperiodic random membrane potential activity. Combining the electrophysiological simulation of simple neural sphere under full initial conditions and hysteresis complex iteration theory of simplified neural sphere dynamical system, the fundamental principle of generating different potential activities in neural sphere is revealed. Based on the information features of neural sphere, we proposed a possible spatio-temporal encoding high-performance neuromorphic computing architecture of brain for the memory and processing of biological information. It can explain the unique functionalities of the brain such as associated information, multiple storage, low power precision operation, computing-in-memory, information compression, feature extraction etc. Based on the estimation by the developed architecture, it is surprisingly discovered that the storage capacity and computational efficiency of the human brain are not only significantly better than the artificial silicon-based systems, but also 3 orders higher than the original estimation of the performance of the brain. Furthermore, based on Landauer's principle, this estimate is reasonable and shows that the brain is extremely energy-efficient.

**Neural sphere for simulating the brain**

In the cerebral cortex, a large number of local neural connections can realize complex functions such as visual perception and emotional processing. These sets of biological neural connections with specific functions are defined as neural ensembles (NEs). To explore the working mechanism of the large and complex network of the brain, it is necessary to first focus on the behavior of the local NEs. In this work, a neural sphere (NS) that unifies the morphological and functional of the brain is developed as shown in Fig. 1 to simulate the electrophysiological activity of the NE. It includes the construction of digital neurons and the optimization of the distribution of neural connections. The detailed constructions, optimization of the NS and simulation process are presented in **Materials and Methods** of Eq. (S.1)-(S.33), Movie S1 and Table S1-S2.

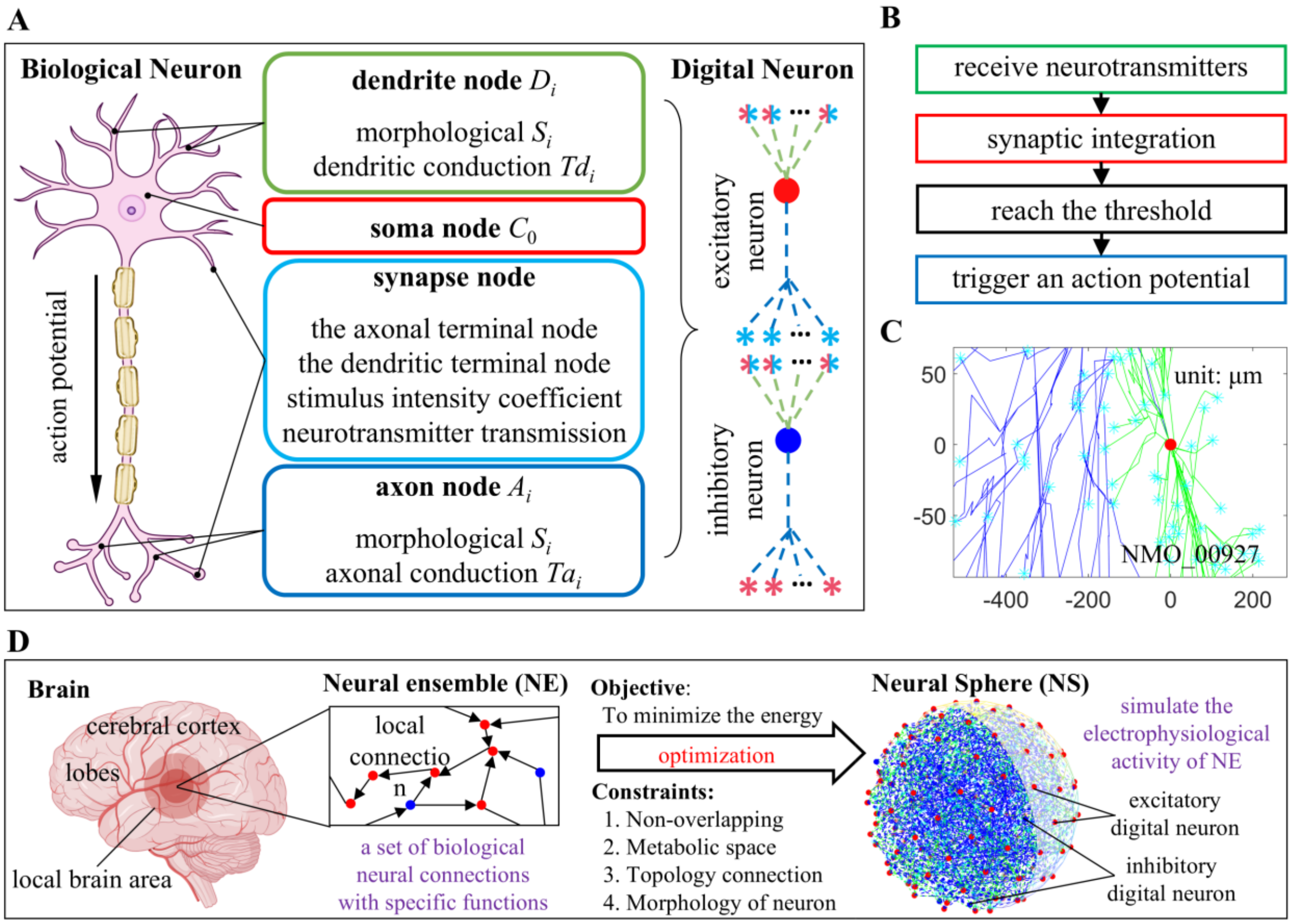

**Fig. 1. Schematic of the neural sphere (NS).** (**A**) Construction of the digital neuron. It consists of four types of nodes, including dendrite node, soma node, axon node and synapse node. The right side of this subfigure shows the schematic representation of the excitatory and inhibitory digital neurons. The red dot indicates the soma of excitatory neuron. The blue dot indicates the soma of inhibitory neuron. The green dashed lines represent the dendrites. The dashed blue lines represent the axons. The cyan asterisks indicate the excitatory synapses. The red asterisks indicate the inhibitory synapses. (**B**) Flowchart of input-output for digital neuron implementation. (**C**) Example of NMO_00927 excitatory neuron from Digitally Reconstructed Neurons and Glia17 plotted through digital neuron. (**D**) Optimization and construction of the NS. The NS is a sphere structure connected by multiple digital neurons that conforms to the minimum energy consumption criterion

In order to describe the microstructure and electrophysiological function(*6-10, 27*), the biological neuron need to be parameterized into the digital neuron by a group of orderly connected nodes as shown in Fig. 1A. Each node includes the morphological information of the neuron extracted by the neuron digital reconstruction technology(*17*) and the functional information describing the electrophysiology. Depending on the neurotransmitter from the synaptic output, digital neurons are stained as excitatory digital neurons and inhibitory digital neurons. By associating adjacent nodes, the digital neuron can simulate the process of receiving, integrating and transmitting information (Fig. 1B). For example, the NMO_00927 neuron from the Digitally Reconstructed Neurons and Glia(*17*) database can be plotted against digital neuron (Fig. 1C).

Through the connection of digital neurons, the connection topologically equivalent to NE can be restored. In the determined topological connection, the microstructure and spatial position of neurons in NE will affect the energy consumption of NE. With the objective function of minimizing the energy consumption of NE, the optimal position distribution of neurons in NE can be obtained by mathematical optimization as shown in Fig. 1D. Its constraints include: (1) neurons must not overlap, (2) neurons must be within maximum metabolic space, (3) neural topological connections remain unchanged, and (4) axons and dendrites of each neuron remain in a morphology with minimum energy consumption. It is found the sphere distribution of neuronal soma positions will make the NE of symmetric topology connection consumes the least energy. In the NE of the general topology connection, the sphere distribution is expected to stably represent at least a suboptimal solution. Based on this conclusion of optimization, a set of digital neurons with ordered connection and sphere distribution conforming to the minimum energy consumption criterion is defined as NS as shown in Fig. 1D. The somas are evenly distributed on the surface of the sphere following the Fibonacci grid(*28*), and the connections between digital neurons are distributed inside the sphere. NS can efficiently simulate membrane potential activity of the NE. After setting the initial membrane potential state of each digital neuron, the NS can simulate the action potential activity of neurons without interruption. In simulations of small-scale (< 1000 neurons) NSs, only synaptic activity inside the NS is considered, while synapses connected to the outside of the ensemble remain silent. The parameters of digital neurons adopt the statistics of real neurons from the neurons digital reconstruction technique(*17*).

## Connection probability & initial condition affect the electrophysiological activity of NS

Through a series of systematic simulations of small-scale NSs, it has been found that the characteristics of electrophysiological activity are significantly affected by the connection probability *p* and initial condition. Taking the NS of 10 digital neurons as an example, Figure 2 shows the membrane potential activities of one of the neurons in the NS. Comprehensive membrane potential activity data with different connection probabilities are presented in Fig. S5-S8.

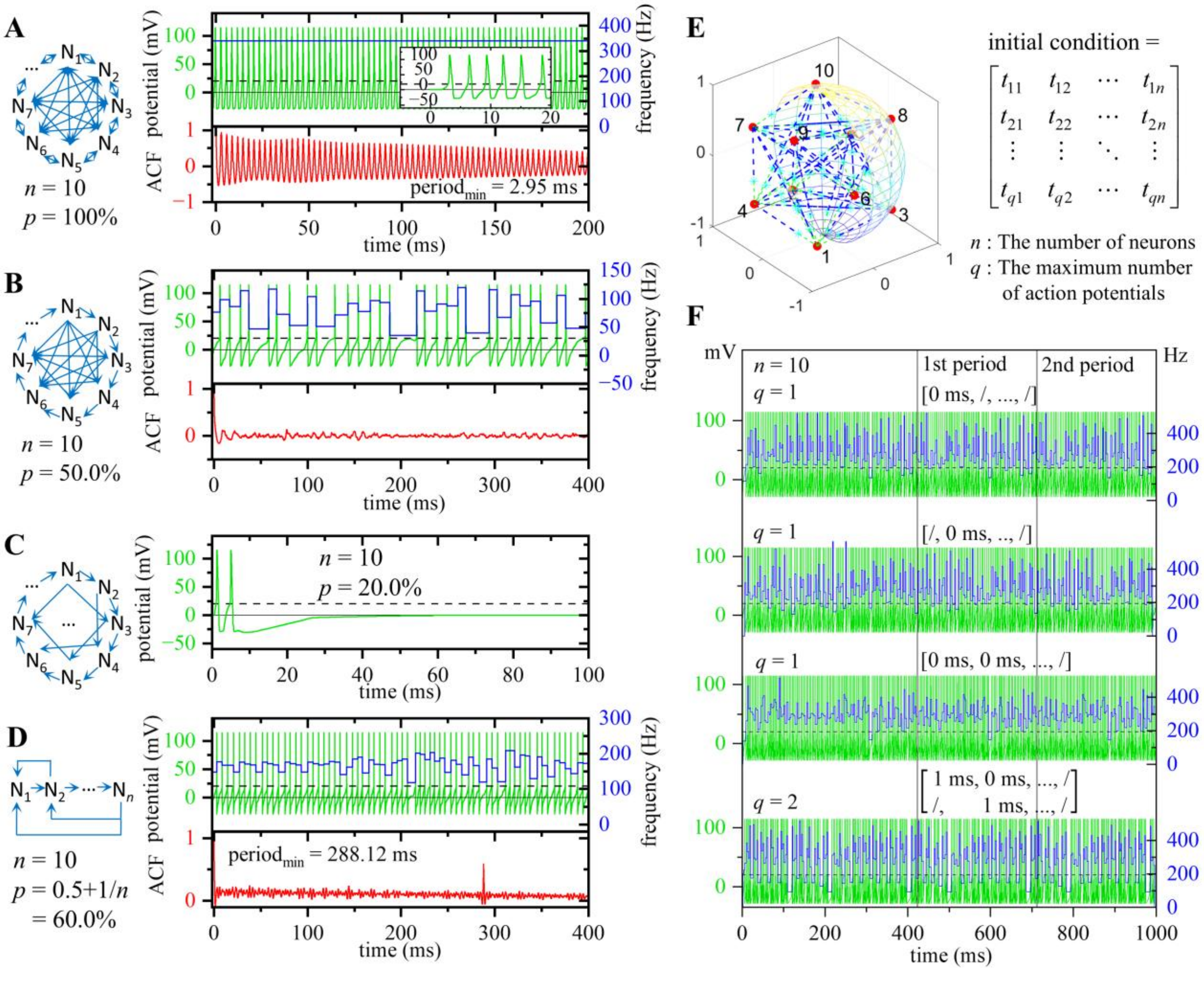

**Fig. 2. The membrane potential activity of NS.** (**A**) Bidirectional fully connected NS. (**B**) Unidirectional fully connected NS. (**C**) Sparsely connected NS. (**D**) Full positively feedback connected NS. (Upper left) Planar schematic of the connection types. Arrows are synaptic connections from axons to dendrites. $N_i$ is the digital neuron with index *i*. (Lower left) The number of digital neurons *n*, and the probability of the connection *p*. (Right) The solid green line is the membrane potential time series of one of the digital neurons. The black dashed line is the threshold potential, 20 mV. The blue solid line is the real-time frequency spectra of action potential as a function of time. The red solid line is the curve of the autocorrelation coefficient function, ACF. (**E**) The matrix representation of the initial conditions of NS. (**F**) Membrane

potential time series and real-time frequency of triggered action potentials for one of the neurons of full positively feedback connected NS of 10 neurons under four initial conditions. The other data are presented in Fig. S9-S18

NS with different connection probabilities exhibit four typical membrane potential activities. In the bidirectional fully connected NS ($p$ = 100%, Fig. 2A), digital neurons exhibit high-frequency action potential firing, approximately 340 Hz. According to the first peak of the autocorrelation coefficient function(*29*), the minimum period of action potential activity is 2.95 ms. This frequency is in the “ultra-fast” oscillatory band (200-600 Hz)(*30*) of the mammalian cerebral cortex. Local NEs with high probability of connection will be suitable for rapid responses with repeated high frequencies. In the unidirectional fully connected NS ($p$ = 50%, Fig. 2B), the average and maximum action potential frequency are 76 Hz and 120 Hz, respectively. The action potential activity of this NS is in the oscillatory band of the gamma wave(*30*). It does not have obvious periodicity or has an ultra-long period that exceeds the maximum simulation time. This random action potential activity implies that NEs will generate self-organizing information to support information processing and cognitive functions. In the sparsely connected NS ($p$ = 20%, Fig. 2C), the digital neurons return to resting state after triggering a few action potentials. It is suitable for structurally simple local NE that provides transient reflections. In the full positively feedback connected NS (Fig. 2D), the action potential activity has an ultra-long period, reaching 288.11 ms, which is on the same level as the average human reaction time(*31*). The average and maximum action potential frequency are 167 Hz and 209 Hz, respectively, belonging to high-frequency gamma waves(*30*). It means that a large amount of ordered information is able to be stored in the NE through periodic changes in action potential activity to realize the learning and memory functions of the brain.

Based on the comprehensive numerical experiments of different $p$, it is found that most of the NSs within $40\% < p < 70\%$ have long period (T > 100 ms) membrane potential activity. The set of periodic membrane potential activity of all neurons in the NS is defined as the periodic potential waveform. These periodic potential waveforms are significantly affected by the initial conditions. The initial condition for the NS with $n$ neurons can be represented as a matrix with $q$ rows and $n$ columns (Fig. 2E). $q$ represents the maximum number of action potentials triggered. $t_{ji}$ represents neuron $i$ triggering $j$-th action potential at moment $t_{ji}$. “$t_{ji}$ = /” represents no action potential triggered. Figure 2F shows the periodic membrane potential time series and the real-time frequency of triggered action potentials for one of the neurons of the full positively feedback connected NS under four initial conditions. In these examples, different initial conditions will converge to different periodic membrane potential activities through NS.

### Information features of NS

The four typical potential activities of NS preliminarily explain the underlying functions of the bio-nervous system, such as unconditioned reflexes, information processing, conditioned reflexes, memory and so on (Fig. 2). Corresponding to all sets of the initial condition matrix, the universal set of potential activities describes the information features of NS. By combining electrophysiological simulation with the analysis of chaos theory for the NS dynamical system (Fig. 3), the fundamental principle of the different types of potential activities exhibited by NS can be well revealed. Detailed simulation data and theoretical modeling of the dynamical system are presented in Data S1 and **Materials and Methods** of Eq. (S.34)-(S.39).

In order to characterize the information features of NS, 221551 groups of electrophysiological activities under all initial conditions of the full positively feedback NS with $n = 5$ and $q = 1$ are simulated. These initial conditions cover all situations that guarantee that the NS generates sustained potential activity. By calculating the first peak of the autocorrelation coefficient function of the membrane potential activity time series for all electrophysiological simulations, two frequency histograms of roughly lognormal distribution in Fig. 3A and Fig. 3B are obtained. The autocorrelation coefficient close to 1 or 0 indicates that the potential activity has stronger periodicity or chaotic property, respectively. According to the cumulative probability curve in Fig. 3C, most of the potential activity (94%) will converge to period potential activity. Only a few potential activities are random potential activities (1%) and periodic potential activities (5%). Besides, neurons located in the middle of the full positively feedback NS are more capable of generating random potential activity (Fig. 3B).

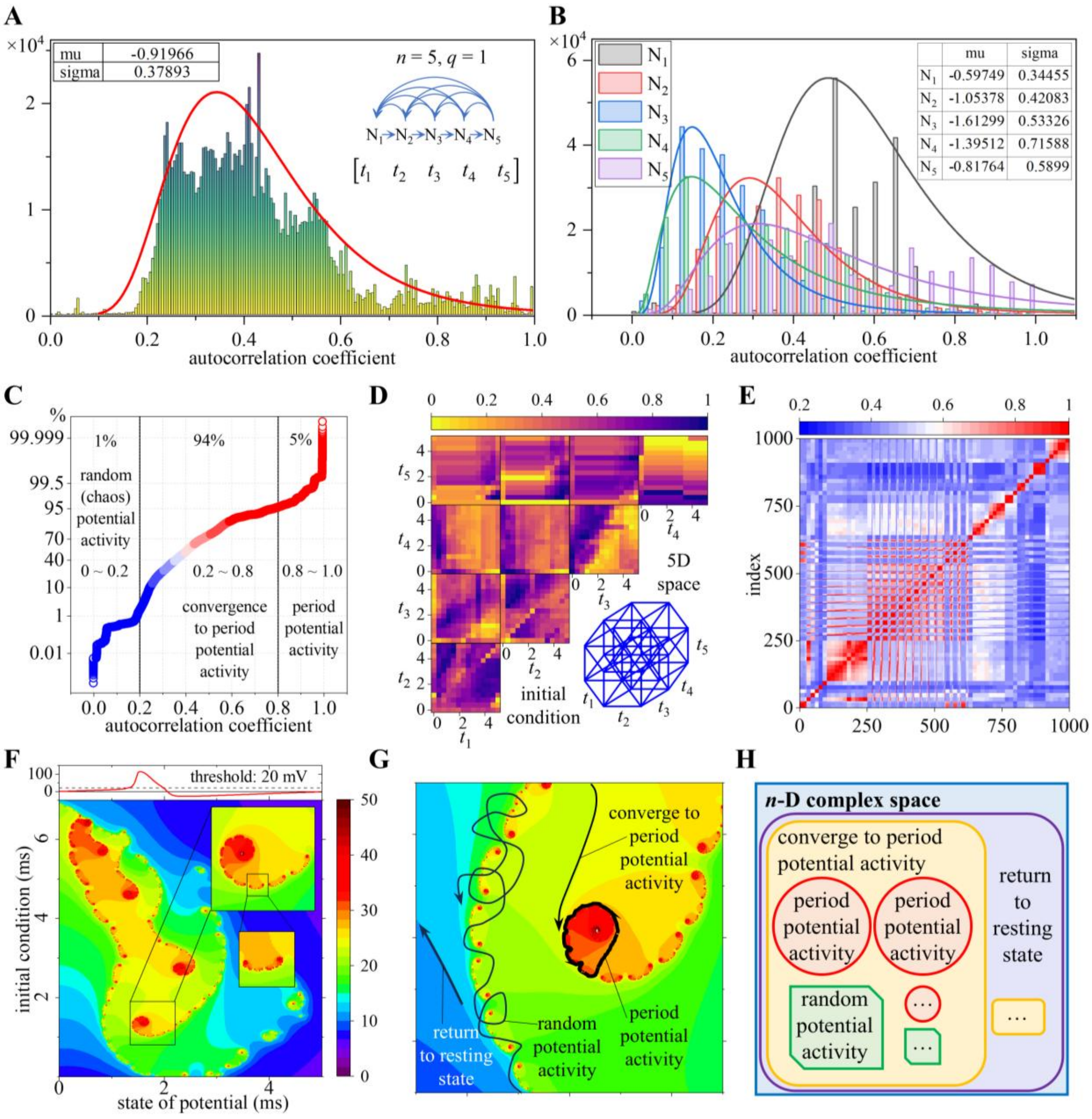

**Fig. 3. Electrophysiological simulation of NS and hysteresis complex iteration theory of simplified NS dynamical system.** (**A**) The frequency histogram of the autocorrelation coefficient of all neuronal potential activities under all initial conditions. The simulation object is the full positively feedback NS of five neurons ($n = 5$). The initial condition is that each neuron triggers an action potential at most once ($q = 1$). The connection mode of NS and the initial condition matrix are given in the upper right corner of the figure. Subfigures A-E all adopt above simulation data. (**B**) The frequency histogram of the autocorrelation coefficient of the potential activity of each neuron under all initial conditions. (**C**) The cumulative probability curve of the autocorrelation coefficient of all neuronal potential activities under all initial conditions. The three intervals of the autocorrelation coefficient respectively represent three typical potential activities. (**D**) All planar projections of the autocorrelation coefficient heat map in the five-dimensional space spanned by the initial condition matrix. (**E**) Heat map of potential activity correlation coefficients under 1000 similar initial conditions. (**F**) Phase portrait of one of the neurons for simplified NS dynamical system. The potential state on the horizontal coordinate

corresponds to the action potential state at each moment. The time of hysteresis on the vertical coordinate corresponds to the initial condition $t_i$. The color graph represents the number of iterations. Other data are presented in Fig. S19-S23 and Movie S2. (**G**) Schematic diagram of the trajectories of four typical potential activities in phase portrait. (**H**) Schematic diagram of the positions and containment relationships of four typical potential activities in the *n*-dimensional complex space spanned by the initial condition and potential state.

In the five-dimensional space spanned by the initial conditions, similar autocorrelation coefficients exhibit distinct clustering characteristics on ten plane projections (Fig. 3D). It indicates that the initial conditions for the same type of potential activity are similar. Besides, by calculating the correlation coefficient of the potential activities generated by 1000 groups of similar initial conditions from the simulation data (Fig. 3E), it is found that most of the potential activities have weak correlation with each other (< 0.5). It means that NS can well distinguish initial conditions with slight differences.

In order to explain the above mathematical principle of the information features of NS, a simplified NS dynamical system is constructed to analyze how the initial conditions drive the potential activity of NS. The state of each neuron is described by a complex number with the real part representing the state of action potential and the imaginary part representing the initial condition $t_i$. Each complex iteration of the dynamical system is equivalent to the synaptic integration of each neuron for several hysteresis inputs from the remaining connected neurons. In the full positively feedback NS dynamical system with $n = 5$ and $q = 1$, staining the complex plane of one of the neurons according to the number of iterations, a phase portrait with a beautiful fractal structure and a large number of strange attractors can be obtained as shown in Fig. 3F. As shown in Fig. 3G, an attractor corresponds to a period potential activity. The point of convergence to the attractor corresponds to convergence to period potential activity. Randomly being alternately attracted by several attractors corresponds to random potential activity. The proportion of area in the phase portrait of the above three cases is consistent well with the simulated data in Fig. 3C. Taking the initial state of each neuron as a dimension, the NS dynamical system will constitute an *n*-dimensional complex space (Fig. 3H). The periodic potential waveform is a high-dimensional strange attractor in this space. Without limiting the time resolution, there will be an infinite number of strange attractors in the NS dynamical system. To sum up, this simplified NS dynamical system is in good agreement with the simulation results, preliminary characterization of the information features of the NS.

## High-performance neuromorphic computing architecture of brain

According to the information features of the NS, a high-performance neuromorphic computing architecture of brain (HNCA) is constructed to describe a possible working mode of the brain (Fig. 4). Different from traditional silicon-based architectures and peak-based neuromorphic computing(*4*), this non von Neuman architecture has remarkable advantages in terms of energy efficiency, storage capacity, spatial footprint, and computing speed.

The HNCA is mainly composed of two kinds of NS, including processing NS of memory NS. The processing NS exhibits the characteristic of random potential activity (box II in Fig. 4A). It can directly receive a string of multisensory stimulus or processed external information and generate self-organized information through aperiodic action potential activity. The memory NS exhibit the characteristic of long period potential activity (box III in Fig. 4A). Through the regulation of short-term synaptic plasticity from a group of neurons, the input to the memory NS is a group of action potentials that trigger sequentially at specific moments (box I in Fig. 4A). It is equivalent to the group of membrane potential states of all neurons at a particular instant in the periodic potential waveform (the red dashed line of Fig. 4B). The set of initial conditions to the set of periodic potential waveforms is a non-injective surjection. Through the set of input information from outside the NS forms a one-to-one mapping with the set of initial conditions (Set A to Set B in Fig. 4C), the input information can be stored in the NS (Set A to Set C in Fig. 4C).

The information processed or stored by NS can be directly output or reprocessed (Box IV in Fig. 4). (1) Through the integration of small-scale NS with different structures and functions, the biological information flow will be iteratively processed and memorized across different NS. In particular, the integration of NSs can be designed as a multi-layer fractal structure. The integration comprising one layer of fractal structures, which includes $m'$ sub NSs each containing $m$ neurons, is defined as fractal of $m' \times m$. And so on, the integration comprising $n$ layer of fractal structures can be represented as fractal of $m^{(n)} \times m^{(n-1)} \times \ldots \times m'' \times m' \times m$. The box IV.1 in Fig. 4A shows that 10 sub NSs of 10 neurons form a NS of 100 neurons through one-layer of fractals, i.e., fractal of 10’×10. (2) By re-inputting the external stimulus, the memory NS will produce a periodic potential waveform to output the memory. (3) The NS can be precompiled to have specific physiological functions through long-term evolution. A group of effectors connected to it can output unconsciousness physiological activity such as the respiratory system, the digestive system, some conditioned reflexes, etc.

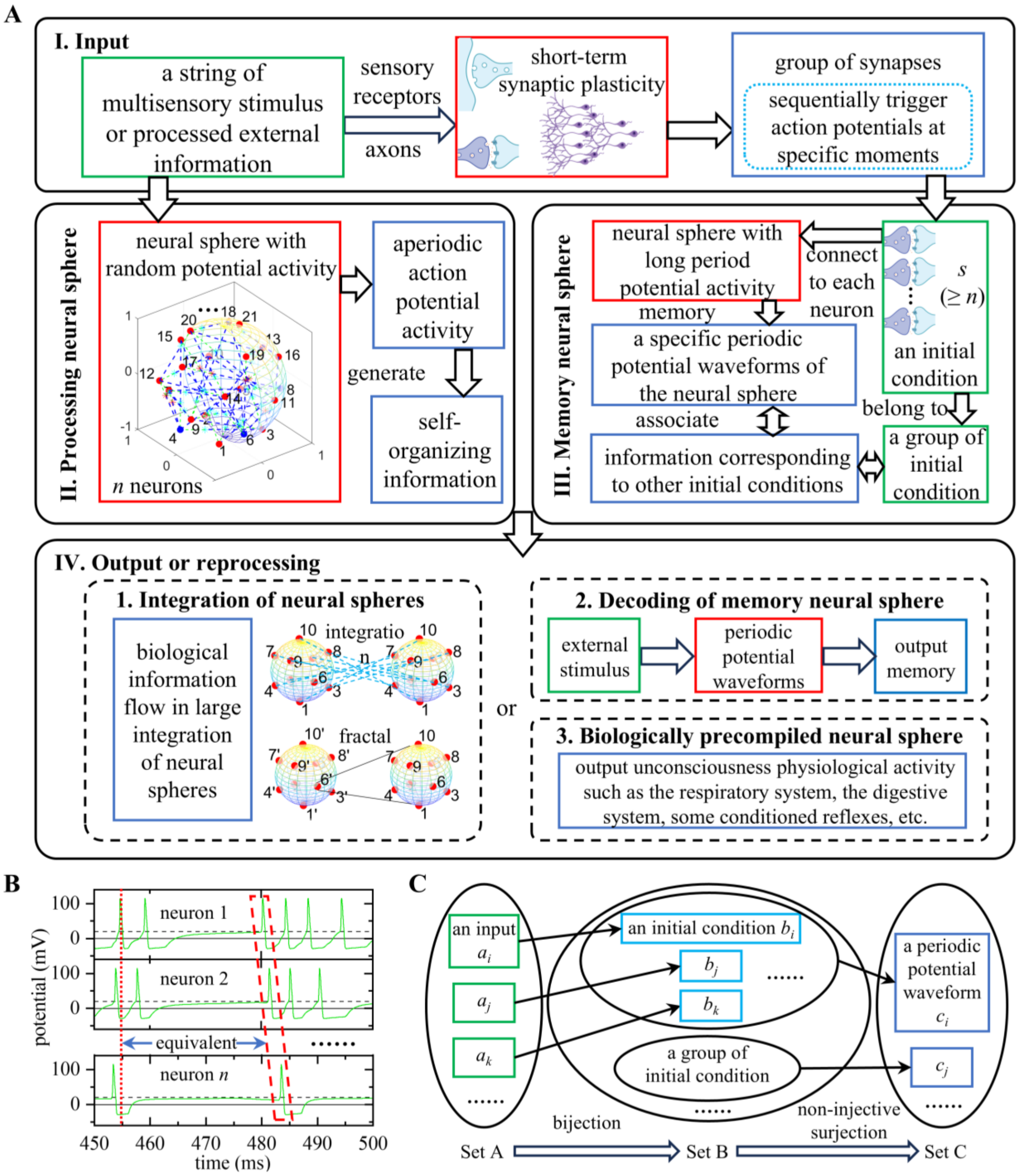

**Fig. 4. High-performance neuromorphic computing architecture of brain (HNCA).** (**A**) The operation mechanism of HNCA. The black rounded boxes show the four modules. Each module can be independent or integrated with each other. The green, red and blue boxes represent the input driver, processing logic and output driver, respectively. In box III, $s$ is the number of synapses. In box IV.1, the cyan dashed lines represent left-to-right or right-to-left synaptic connections between neurons. (**B**) Schematic diagram of two equivalent initial conditions. (**C**) Mapping relationship from input to periodic potential waveforms of NS

## HNCA exhibits super storage capacity and computational power

In order to show the advantages of HNCA, it is necessary to quantify its storage capacity and computational power through the principles of Shannon information theory(*32*) and floating point operations per second (FLOPS)(*33*), respectively. The detailed calculations are provided in **Materials and Methods** of Eq. (S.40)-(S.55).

By calculating the self-information of several artificial systems and HNCAs, the curves of the self-information and its slope as a function of the number of devices or neurons are obtained in Fig. 5A and Fig. 5B, respectively. The binary system, the architecture of current computers, has the lowest self-information. Its slope is 1 bit per transistor. According to Fig. 4C, the self-information of HNCA is obtained on the basis of estimating the number of periodic potential waveforms. It is closely related to factors such as the temporal resolution, the scale of the NS, the connectivity of the synapses, etc. When the time resolution is the physical limit (Planck time), the self-information of small-scale HNCA is 2 or more orders of magnitude larger than that of the binary systems. The self-information of the integration of fractal of $m'\times m$ ($m$ = 5, 6, …, 10) is slightly larger than that of the non-integrated HNCA. This benefits from the fact that the fractal structure reprocesses the periodic potential waveforms of sub memory NSs to obtain additional self-information. The conservative and aggressive estimates of the storage capacity for the human brain modeled by HNCA are $8.24\times10^{15}$ Bytes and $7.48\times10^{18}$ Bytes, respectively. It is much larger than the previous maximum estimate of $1\times10^{15}$ Bytes.(*26*)

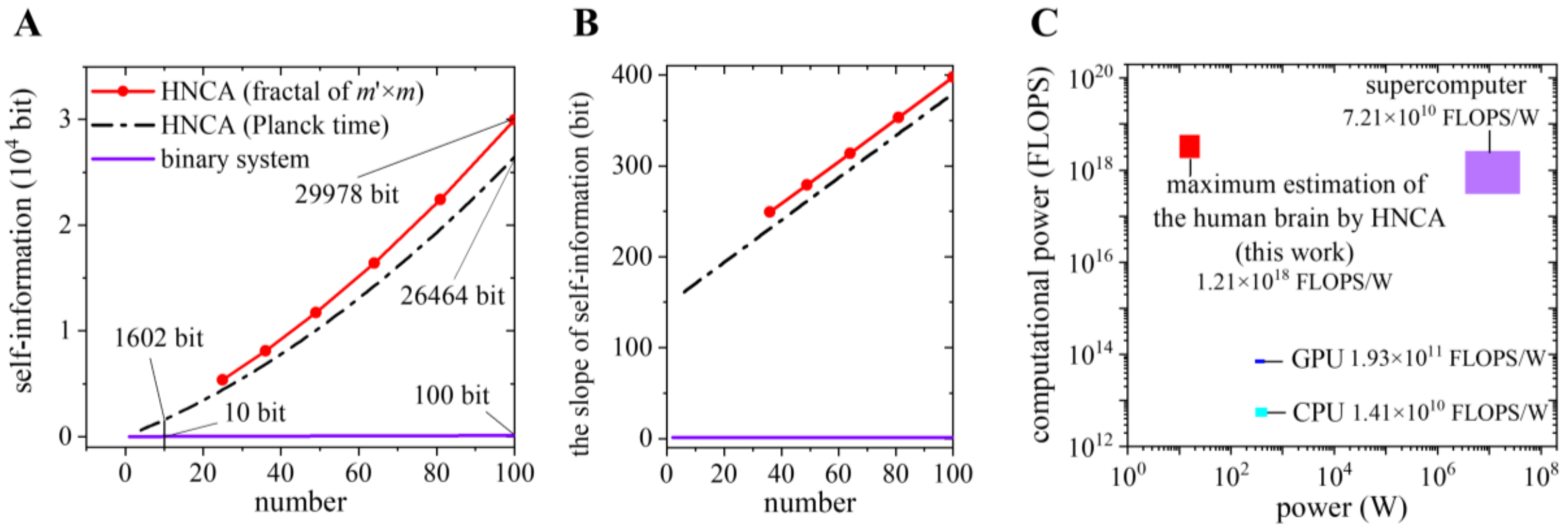

**Fig. 5. Evaluation of storage capacity and computational efficiency for HNCA and the binary system.** (**A**) Curves of the self-information of different systems as a function of the number of neurons or devices. The black line represents the HNCA with a Planck time (the physical limit of time, $5.39\times10^{-44}$ s) as its temporal resolution. The red line represents six HNCAs with the fractal of $m'\times m$ ($m$ = 5, 6, …, 10). More detailed comparison data are in Fig. S24. (**B**) The slopes of the self-information curves in **A**. (**C**) Comparison of computational efficiency between the human brain estimated by HNCA and silicon-based devices. The computing power is measured by double precision floating point operations per second. The average computational efficiency of each system (FLOPS/W) is shown in the figure. The source of the original data is in Table S4.

Although HNCA is not well-suited for floating-point operations, the integrated output of a single action potential or the memory space operable per unit time can be equivalent to several floating-point operations to measure its computational power. Through the above two equivalent methods, the maximum computational power of the human brain modeled by HNCA is estimated to be $1.72\times10^{18}$ FLOPS and $6.24\times10^{18}$ FLOPS at most. (Fig. 5C) Its computational power is slightly higher than that of the supercomputer and is equivalent to 78,000 latest generation GPUs. Since the power of the human brain is only about 20 W, its average computational efficiency is 7 to 8 orders of magnitude higher than that of the silicon-based chip.

Based on Landauer's principle(*34, 35*), it is necessary to analyze the physical limit of the power for HNCA. According to the maximum estimate in Fig. 5C, the Landauer limit of the human brain simulated by HNCA is about 2.37 W according to Eq. (S.57). Based on the existing experimental data, the power of the human brain for computing and communication is about 3 W(*36*), which is only 1.26 times the Landauer limit. Its energy efficiency has reached 79%. However, modern computers are a billion times beyond their Landauer limit, far less energy-efficient than biological systems such as the human brain. Even the most advanced chips at present have an energy efficiency of only about $1\times10^{-7}$%. It means that the energy efficiency of the human brain has almost reached its limit after a long period of evolution. In other words, it also verifies the rationality of HNCA in the estimated storage capacity and computing power.

## Discussion

According to the sphere distribution of NE with optimal energy consumption, the neural sphere (NS) is constructed, which integrates the morphology of microscopic neurons, the topological connections of macroscopic NEs, and the electrophysiological mechanisms at various scales (Fig. 1). Based on the potential activity characteristics of NS, we propose a high-performance neuromorphic computing architecture of brain (HNCA) to describe the possible working mechanism of brain (Fig. 4), initially explaining the characteristics of large capacity and high computational efficiency of brain.

The HNCA can be extended from small-scale ensembles to large-scale ensembles, or even the whole brain, through ingenious integration. According to the simulations of the NS, the more complex connections are difficult to regulate to ensure that it has stable periodic characteristics as the scale of NS increases. The large-scale NSs with ultra-long period will greatly increase the upper limit of time for reading information. By integrating in series and parallel, large-scale NSs are able to control the stability of periodic properties and the upper limit of time for reading information within the level of a sub NS.

Compared with the binary architecture of current computer, HNCA has more significant advantages. Firstly, HNCA can associate different inputs. This feature is closely related to phenomena such as associative memory, jumping thinking and so on. Secondly, HNCA has the function of multiple storage. A memory NS has a large number of different periodic potential waveforms to store a large number of independent information groups simultaneously. It has better anti-interference between different information groups. Thirdly, in order to realize the precise operation of the brain with low power consumption, HNCA has the function of offline storage. Once the storage or reading of information is completed, inhibitory neurons will be triggered to inhibit the activity of the memory NS. Fourthly, through the coupling of memory NSs, stored information of HNCA can be reprocessed to achieve integrated storage and computation functionality, i.e., computing-in-memory. Fifthly, HNCA can compress the huge initial condition information to extract the feature information for storage.

In order to further develop the HNCA, it still needs a lot of additional technical and theoretical supports. The simulation of neural network action potential activity requires more computational power and more accurate time resolution at biological limits. On the basis of accurate simulation, it is necessary to predesign a NS with ultra-long periodic action potential activity. Besides, the brain needs a new quantitative method of bio-information and bio-computational power instead of Shannon information theory and FLOPS, respectively.

With the rapid advancement of AI Generated Content technology, the increasing demand for computational power and energy efficiency is driving the emergence of higher energy-efficient computing architectures, including neuromorphic computing architectures, computing-in-memory architectures, etc. Due to the limitations of the size and power consumption of field-effect transistors, the improvement in energy efficiency of current computing architectures has almost reached a bottleneck. Therefore, exploring the information principles of the brain consisting of micron scale neurons will provide a possible path to achieving Artificial General Intelligence. We hope that the results and insights of this work will serve as an important reference for explaining the efficient working of the brain to facilitate detailed investigation of information mechanisms in biological neural networks and inspire novel biomimetic computing architectures.

## References and Notes

1. D. Silver *et al.*, Mastering the game of Go with deep neural networks and tree search. *Nature* **529**, 484-489 (2016).
2. H. C, G. P, A. D, Updated Energy Budgets for Neural Computation in the Neocortex and Cerebellum. *Journal of Cerebral Blood Flow & Metabolism* **32**, 1222-1232 (2012).
3. K. Roy, A. Jaiswal, P. Panda, Towards spike-based machine intelligence with neuromorphic computing. *Nature* **575**, 607-617 (2019).
4. D. Kudithipudi *et al.*, Neuromorphic computing at scale. *Nature* **637**, 801-812 (2025).
5. S. Cajal, Estructura de los centros nerviosos de las aves. *Rev. Trim. Histol. Norm. Patol.* **1**, 1-10 (1888).
6. A. L. Hodgkin, A. F. Huxley, B. Katz, Measurement of current-voltage relations in the membrane of the giant axon of loligo. *The Journal of Physiology* **116**, 449-472 (1952).
7. A. L. Hodgkin, A. F. Huxley, Currents carried by sodium and potassium ions through the membrane of the giant axon of loligo. *The Journal of Physiology* **116**, 449-472 (1952).
8. A. L. Hodgkin, A. F. Huxley, The components of membrane conductance in the giant axon of loligo. *The Journal of Physiology* **116**, 473-496 (1952).
9. A. L. Hodgkin, A. F. Huxley, The dual effect of membrane potential on sodium conductance in the giant axon of loligo. *The Journal of Physiology* **116**, 497-506 (1952).
10. A. L. Hodgkin, A. F. Huxley, A quantitative description of membrane current and its application to conduction and excitation in nerve. *The Journal of Physiology* **116**, 500-544 (1952).
11. R. D. Traub, R. K. Wong, R. Miles, H. Michelson, A model of a CA3 hippocampal pyramidal neuron incorporating voltage-clamp data on intrinsic conductances. *Journal of neurophysiology* **66**, 635-650 (1991).
12. X. J. Wang, G. Buzsáki, Gamma oscillation by synaptic inhibition in a hippocampal interneuronal network model. *Journal of neuroscience* **16**, 6402-6413 (1996).
13. P. J. Basser, J. Mattiello, D. LeBihan, MRI of neuronal network structure, function, and plasticity. *Biophysical Journal* **66**, 259-267 (1994).
14. N. A. Steinmetz *et al.*, Neuropixels 2.0: A miniaturized high-density probe for stable, long-term brain recordings. *Science* **372**, eabf4588 (2021).
15. N. J. Sofroniew, D. Flickinger, J. King, K. Svoboda, A large field of view two-photon mesoscope with subcellular resolution for in vivo imaging. *eLife* **5**, e14472 (2016).
16. C. Stringer, M. Pachitariu, Analysis methods for large-scale neuronal recordings. *Science* **386**, eadp7429 (2024).
17. G. A. Ascoli, D. E. Donohue, M. Halavi, NeuroMorpho.Org: A Central Resource for Neuronal Morphologies. *Journal of Neuroscience* **27**, 9247-9251 (2007).

18. J. G. White, E. Southgate, J. N. Thomson, S. Brenner, The structure of the nervous system of the nematode Caenorhabditis elegans. *Philos Trans R Soc Lond B Biol Sci* **314**, 1-340 (1986).
19. A. Lin *et al.*, Network statistics of the whole-brain connectome of Drosophila. *Nature* **634**, 153-165 (2024).
20. N. L. Turner *et al.*, Reconstruction of neocortex: Organelles, compartments, cells, circuits, and activity. *Cell* **185**, 1082-1100 (2022).
21. C. R. Gamlin *et al.*, Connectomics of predicted Sst transcriptomic types in mouse visual cortex. *Nature* **640**, 497-505 (2025).
22. O. Sporns, G. Tononi, R. Kötter, The human connectome: a structural description of the human brain. *PLoS Comput. Biol.* **1**, e42 (2005).
23. M. Naddaf, Europe spent €600 million to recreate the human brain in a computer. How did it go? *Nature* **620**, 718-720 (2023).
24. O. Sporns, R. Kötter, Motifs in brain networks. *PLOS Biol.* **2**, e369 (2004).
25. L. Luo, Architectures of neuronal circuits. *Science* **373**, eabg7285 (2021).
26. T. M. B. Jr *et al.*, Nanoconnectomic upper bound on the variability of synaptic plasticity. *eLife* **4**, e10778 (2015).
27. W. Rall, Branching dendritic trees and motoneuron membrane resistivity. *Experimental neurology* **1**, 491-527 (1959).
28. Gonzalez, Alvaro, Measurement of Areas on a Sphere Using Fibonacci and Latitude-Longitude Lattices. *Mathematical Geosciences* **42**, 49-64 (2010).
29. K. I. Park, M. Park, in *Fundamentals of probability and stochastic processes with applications to communications*. (Springer, Cham, 2018), pp. 165-168.
30. G. Buzsáki, A. Draguhn, Neuronal Oscillations in Cortical Networks. *Science* **304**, 1926-1929 (2004).
31. R. J. Kosinski, A literature review on reaction time. *Clemson University* **10**, 337-344 (2008).
32. C. E. Shannon, A mathematical theory of communication. *The Bell System Technical Journal* **27**, 379-423 (1948).
33. K. D. J, *Parameters of a Stochastic Process*. Computer system capacity fundamentals (National Bureau of Standards, US Department of Commerce, 1974).
34. A. Bérut *et al.*, Experimental verification of Landauer's principle linking information and thermodynamics. *Nature* **483**, 187-189 (2012).
35. Y. Jun, M. Gavrilov, J. Bechhoefer, High-Precision Test of Landauer's Principle in a Feedback Trap. *Phys. Rev. Lett.* **113**, 190601 (2014).
36. P. Lennie, The Cost of Cortical Computation. *Current biology* **13**, 493-497 (2003).
37. Z. Yao *et al.*, A high-resolution transcriptomic and spatial atlas of cell types in the whole mouse brain. *Nature* **624**, 317-332 (2023).
38. S. J. Cook *et al.*, Whole-animal connectomes of both Caenorhabditis elegans sexes. *Nature* **571**, 63-71 (2019).
39. Z. Zheng *et al.*, Unconventional ferroelectricity in moiré heterostructures. *Nature* **588**, 71-76 (2020).
40. X. W. K. Yasuda, K. Watanabe, T. Taniguchi, P. Jarillo-Herrero, Stacking-engineered ferroelectricity in bilayer boron nitride. *Science* **372**, 1458-1462 (2021).
41. Q. Li *et al.*, A single-cell transcriptomic atlas tracking the neural basis of division of labour in an ant superorganism. *Nature ecology & evolution* **6**, 1191-1204 (2022).
42. D. Sharma *et al.*, Linear symmetric self-selecting 14-bit kinetic molecular memristors. *Nature* **633**, 560-566 (2024).

**Acknowledgments:** We thank Xiaofan Wang, Xiaokai Chen, and Xiaoyu Xuan for their discussions.

**Funding:** This work was supported by the National Natural Science Foundation of China (T2293691), Natural Science Foundation of Jiangsu Province (BK20243065), the Research Fund of State Key Laboratory of Mechanics and Control for Aerospace Structures (MCAS-I-0525K01), the Fundamental Research Funds for the Central Universities (NJ2024001, NC2023001, NJ2023002) and the Fund of Prospective Layout of Scientific Research for NUAA (Nanjing University of Aeronautics and Astronautics).

**Author contributions:** Jinxuan Ma designed the conceptualization and methodology. Jinxuan Ma and Wanlin Guo designed the numerical experiments together. Wanlin Guo has obtained financial support. Wanlin Guo and Jinxuan Ma contributed to the discussion and revision of the manuscript.

**Competing interests:** Authors declare that they have no competing interests.

**Data and materials availability:** The neuron morphology data are obtained from Digitally Reconstructed Neurons and Glia, https://neuromorpho.org/. The data of the latest generation high-performance CPU and GPU and the top 10 supercomputer are obtained from various institutions or manufacturers, including https://www.intel.com/content/www/us/en/ark.html#@Processors, https://www.amd.com/en/products/processors/server/epyc/9005-series.html, https://www.amd.com/en/products/graphics/workstations.html, https://www.nvidia.com/en-us/geforce/graphics-cards/40-series/rtx-4090/, and https://www.top500.org/lists/top500/2024/11/. Other datasets collected and/or analyzed during the current study are available from the corresponding author upon request. Source data are provided with this paper in the **Supplementary Materials**. Other data in this study are available from the corresponding author upon request.

**Supplementary Materials**

Materials and Methods

Supplementary Text

Figs. S1 to S25

Tables S1 to S4

Movies S1 to S2

Data S1

References (*37-42*)

# Supplementary Materials for

## High-performance neuromorphic computing architecture of brain

Jinxuan Ma, Wanlin Guo

Corresponding author: Wanlin Guo, wlguo@nuaa.edu.cn

**The PDF file includes:**

Materials and Methods
Supplementary Text
Figs. S1 to S25
Tables S1 to S4
References

**Other Supplementary Materials for this manuscript include the following:**

Movies S1 to S2
Data S1

## Materials and Methods

Construction of the NS

This section provides the detailed method for the construction of the neural sphere (NS), including the construction of digital neuron to describe the structure and function at the microcosmic scale (Fig. S1A-G), the optimization of the sphere distribution for the neural ensemble (NE) (Fig. S2), and the sphere connection structure to describe the synaptic connections of neural networks at the macroscopic scale (Fig. S1h).

(1) Digital neuron

When simulating the membrane potential activity, the design of a single neuron model is essential. The digital neuron is constructed to describe the characteristics of neurons and applies it to neural network simulations (Fig. 1A). The digital neuron takes into account both the functionality and morphology of individual neurons, enabling efficient simulation of interactions between neurons. Figure S1A is an example of NMO_00927 excitatory neuron from Digitally Reconstructed Neurons and Glia(*17*) plotted through digital neuron.

The dendrites and axons of multi-level neurons often exhibit numerous branching structures, forming intricate spatial trees. Neurons can be equivalently represented as a collection of branching or turning points within the spatial trees of dendrites and axons

$$N = \{C_0, D_i, A_j\},\ i = -1, -2, \cdots, -n_\mathrm{d}, j = 1, 2, \cdots, n_\mathrm{a}, \tag{S.1}$$

where $C_0$ represents the soma node, $D_i$ represents the node $-i$ on the dendrite (the numbering of dendritic node takes negative values), $A_j$ represents the node $j$ on the axon (the numbering of axonal node takes positive values), $n_\mathrm{d}$ represents the number of nodes on the dendrite, and $n_\mathrm{a}$ represents the number of nodes on the axon. The terminal nodes of the axon and dendrite are synapse nodes. Each node is a struct that describes the characteristics of neurons at the current location. The following will provide a detailed explanation of how different types of nodes represent the morphology and functionality of neurons, in order the morphological sub struct, axon node, dendrite node, soma node and synapse node.

The morphological sub structs are included in all nodes, describing the position and connection of the current node (Fig. S1B)

$$S_i = (i, j, x_i, y_i, z_i), \tag{S.2}$$

where $i$ is the number of the node, $j$ is the number of the parent node ($|j| < |i|$), and ($x_i$, $y_i$, $z_i$) is the coordinate of the node. In particular, the soma node is represented by (0, -1, 0, 0, 0), and the axon hillock node is represented by (1, 0, 0, 0, 0).

The function of the axon node is to transmit information by action potentials (Fig. S1C). Once an action potential is triggered, it maintains a consistent amplitude as it conducts along the axon. The conduction properties of the axon are represented by quaternion sub structs, referred to as the axonal conduction sub structs

$$Ta_i = (r_{ij}, l_i, v_{ij}, t_i), \tag{S.3}$$

where $r_{ij}$ is the axonal radius between nodes $i$ and $j$. $l_i$ is the path length from node $i$ to the axon hillock node, which is the sum of the path length from node $j$ to the axon hillock node and the path length between nodes $i$ and $j$

$$l_i = l_j + l_{ij} = l_j + \sqrt{(x_i - x_j)^2 + (y_i - y_j)^2 + (z_i - z_j)^2}. \tag{S.4}$$

$v_{ij}$ is the action potential conduction velocity between nodes $i$ and $j$, which is proportional to $r_{ij}^{0.5}$. The conduction velocity of action potentials along the axon can be roughly estimated by $v_{ij} = \alpha r_{ij}^{0.5}$. The velocity coefficient $\alpha$ varies depending on the morphology and function of the neuron and can be determined through simple electrophysiological experiments. $t_i$ represents the time it takes for the action potential to propagate from the axon hillock to node $i$. The calculation formula is given by

$$t_i = t_j + \frac{l_{ij}}{v_{ij}} = t_j + \frac{\sqrt{(x_i - x_j)^2 + (y_i - y_j)^2 + (z_i - z_j)^2}}{v_{ij}} . \tag{S.5}$$

Therefore, the axon node is a nine-dimensional set composed of two sub structs

$$A_i = \{S_i, Ta_i\} = (i, j, x_i, y_i, z_i; r_{ij}, l_i, v_{ij}, t_i),\ i = 1, 2, \cdots, n_{\mathrm{a}}, j = 0, 1, \cdots, n_{\mathrm{a}} - 1 . \tag{S.6}$$

The function of dendrite node is to receive stimuli and aggregate them into the soma (Fig. S1D). The majority of neuronal dendrites are passively conducted. The intensity of the electrical signal on the dendrites gradually decays as the conduction distance increases. After the input of a transient pulse current $I_{\mathrm{e}}$ into the dendritic terminal, the increment of the membrane potential (IMP)(*27*) is

$$V_{\mathrm{m}}(I_{\mathrm{e}}, l, t) = \frac{I_{\mathrm{e}} R_\lambda}{\sqrt{4\pi\lambda^2 t / \tau_{\mathrm{m}}}} \exp(-\frac{\tau_{\mathrm{m}} l^2}{4\lambda^2 t}) \exp(-\frac{t}{\tau_{\mathrm{m}}}) , \tag{S.7}$$

where $t$ is time, $l$ is the distance from the synaptic terminal and $R_\lambda = R_{\mathrm{L}}\lambda / \pi r^2$. The time constant $\tau_{\mathrm{m}} = R_{\mathrm{m}} C_{\mathrm{m}}$ and the electrotonic length $\lambda = (r R_{\mathrm{m}} / 2R_{\mathrm{L}})^{1/2}$ set the scale of the temporal and spatial variation of the membrane potential, where $r$ is radius, $R_{\mathrm{m}}$ is resistance across a unit area of membrane, $R_{\mathrm{L}}$ is specific resistance of the internal medium and $C_{\mathrm{m}}$ is capacitance of membrane.

The conduction properties of dendrites are represented by a ternary sub structs, referred to as the dendritic conduction sub structs

$$Td_i = (r_{ij}, l_i, V_{\mathrm{m}i}) , \tag{S.8}$$

where $r_{ij}$ and $l_i$ are defined the same as in Equation (S.3). $V_{\mathrm{m}i}$ is the IMP at the node $i$

$$V_{\mathrm{m}i} = \sum_{ik} V_{\mathrm{m}ik}(I_{\mathrm{e}ik}(t), l_{ik} - l_i, t) , \tag{S.9}$$

where the subscript $ik$ denotes the $k$-th dendrite ending connected to node $i$, $I_{\mathrm{e}ik}(t)$ represents the current spectrum equivalent to the external stimulus. The IMP at the dendritic root node is represented as

$$V_{\mathrm{m}1}(t) = \sum_{1k} V_{\mathrm{m}1k}(I_{\mathrm{e}1k}(t), l_{1k}, t) . \tag{S.10}$$

Therefore, the dendrite node is an eight-dimensional set composed of two sub structs

$$D_i = \{S_i, Td_i\} = (i, j, x_i, y_i, z_i; r_{ij}, l_i, V_{\mathrm{m}i}),\ i = -1, -2, \cdots, -n_{\mathrm{d}}, j = 0, -1, \cdots, -(n_{\mathrm{d}} - 1) . \tag{S.11}$$

The soma node contains the global eigen parameters of the current neuron, including the coordinates of the soma, threshold potential, the time constant, the electrotonic length, the waveform of the action potential and so on (Fig. S1E).

The synapse node includes the coordinates of the synapse, the axonal terminal node of the presynaptic neuron, the dendritic terminal node of the postsynaptic neuron, the stimulus intensity coefficient, the time of neurotransmitter transmission and so on (Fig. S1F).

Through the cooperation among the above nodes, the neuron also needs to realize the function of receiving and integrating information by synaptic integration. Neurons have numerous connections through synapses to receive input information from other neurons. A chemical synapse transmission requires approximately 0.3 to 0.5 ms. The amplitude of a single postsynaptic potential (PSP) is typically insufficient to trigger an action potential. Multiple synapses must undergo spatial or temporal integration to reach the threshold potential of the neuron (Fig. S1G).

The unit of neurotransmitter release from a synapse is the total amount of neurotransmitter contained in a single vesicle. The PSP is an integer multiple of the unit PSP generated by a single vesicle, meaning that the PSP is quantized. For the sake of analytical simplicity, the peak value of the unit IMP can be set to 0.1 mV. The unit IMP can be represented as

$$V_{\mathrm{m0}}(l,t) = V_{\mathrm{m}}(l,t) = 0.1\exp\left(-\frac{\tau_{\mathrm{m}} l^2}{4\lambda^2 t}\right)\exp\left(-\frac{t}{\tau_{\mathrm{m}}}\right)\ \mathrm{mV}. \qquad \text{(S.12)}$$

When receiving a single stimulus, the IMP can be represented as

$$V_{\mathrm{m}}(t) = k_{\mathrm{V}} V_{\mathrm{m0}}(l,t),\ k_{\mathrm{V}} \in Z, \qquad \text{(S.13)}$$

where $k_{\mathrm{V}}$ is stimulus intensity coefficient (an integer), representing that the IMP is $k_{\mathrm{V}}$ times the unit IMP. When $k_{\mathrm{V}}$ is positive, it represents an excitatory postsynaptic potential (EPSP), and when $k_{\mathrm{V}}$ is negative, it represents an inhibitory postsynaptic potential (IPSP).

The IMP of a neuron can be considered as the spatiotemporal sum of EPSP and IPSP. Assume that neuron $i$ is connected to $j$ axons and continuously receives input from these axons. At time $t_{\mathrm{n}}$, the spatial sum of the external inputs received by the dendritic root node can be represented as

$$V_{\mathrm{m1}}(t) = \sum_{1k} k_{\mathrm{V1}k} V_{\mathrm{m0}}(l_{1k}, t - t_{\mathrm{n}}),\ 1k = 1, 2, \cdots, j, \qquad \text{(S.14)}$$

where $k_{\mathrm{V1}k}$ is the stimulus intensity coefficient of synapse $1k$. When the dendritic root node receives input at time $t_{\mathrm{n}}$, the IMP will be updated as follows to describe the temporal sum of the external inputs

$$V_{\mathrm{m1_new}}(t) = \begin{cases} V_{\mathrm{m1}}(t), & t < t_{\mathrm{n}}, \\ V_{\mathrm{m1}}(t_{\mathrm{n}}) + \sum_{1k} k_{\mathrm{V1}k} V_{\mathrm{m0}}(l_{1k}, t - t_{\mathrm{n}}), & t_{\mathrm{n}} \le t \le t_{\mathrm{n}} + \frac{\tau_{\mathrm{m}} l_0}{2\lambda}, \\ \left(V_{\mathrm{m1}}(t_{\mathrm{n}}) + \sum_{1k} k_{\mathrm{V1}k} V_{\mathrm{m0}}\left(l_{1k}, \frac{\tau_{\mathrm{m}} l_0}{2\lambda}\right)\right) \dfrac{V_{\mathrm{m0}}(l_0, t - t_{\mathrm{n}})}{V_{\mathrm{m0}}\left(l_0, \frac{\tau_{\mathrm{m}} l_0}{2\lambda}\right)}, & t > t_{\mathrm{n}} + \frac{\tau_{\mathrm{m}} l_0}{2\lambda}, \end{cases} \qquad \text{(S.15)}$$

$$V_{\mathrm{m1_new}}(t) = \begin{cases} V_{\mathrm{m1}}(t), & t < t_{\mathrm{n}}, \\ V_{\mathrm{m1}}(t) + \sum_{1k} k_{\mathrm{V1}k} V_{\mathrm{m0}}(l_{1k}, t - t_{\mathrm{n}}), & t_{\mathrm{n}} \le t < t_{\mathrm{n0}} \\ V_{\mathrm{m1}}(t_{n0}) + \sum_{1k} k_{\mathrm{V1}k} V_{\mathrm{m0}}(l_{1k}, t - t_{\mathrm{n}}), & t_{\mathrm{n0}} \le t \le t_{\mathrm{n}} + \frac{\tau_{\mathrm{m}} l_0}{2\lambda}, \\ \left(V_{\mathrm{m1}}(t_{\mathrm{n}}) + \sum_{1k} k_{\mathrm{V1}k} V_{\mathrm{m0}}\left(l_{1k}, \frac{\tau_{\mathrm{m}} l_0}{2\lambda}\right)\right) \dfrac{V_{\mathrm{m0}}(l_0, t - t_{\mathrm{n}})}{V_{\mathrm{m0}}\left(l_0, \frac{\tau_{\mathrm{m}} l_0}{2\lambda}\right)}, & t > t_{\mathrm{n}} + \frac{\tau_{\mathrm{m}} l_0}{2\lambda}, \end{cases} \qquad \text{(S.16)}$$

or

$$V_{\text{m1_new}}(t)=\begin{cases} V_{\text{m1}}(t), & t<t_{\text{n}}, \\ V_{\text{m1}}(t)+\sum\limits_{1k} k_{\text{V1}k}V_{\text{m0}}(l_{1k},t-t_{\text{n}}), & t_{\text{n}}\le t<t_{\text{n}}+\dfrac{\tau_{\text{m}}l_0}{2\lambda} \\ V_{\text{m1}}(t)+\sum\limits_{1k} k_{\text{V1}k}V_{\text{m0}}(l_{1k},\dfrac{\tau_{\text{m}}l_0}{2\lambda}), & t_{\text{n}}+\dfrac{\tau_{\text{m}}l_0}{2\lambda}\le t\le t_{\text{n0}}, \\ (V_{\text{m1}}(t_{\text{n}})+\sum\limits_{1k} k_{\text{V1}k}V_{\text{m0}}(l_{1k},\dfrac{\tau_{\text{m}}l_0}{2\lambda}))\dfrac{V_{\text{m0}}(l_0,t-t_{\text{n}})}{V_{\text{m0}}(l_0,\dfrac{\tau_{\text{m}}l_0}{2\lambda})}, & t>t_{\text{n}}+\dfrac{\tau_{\text{m}}l_0}{2\lambda}, \end{cases} \quad \text{(S.17)}$$

where $t_{\text{n0}}$ is the time at which the neuron's membrane potential reached its peak prior to the update. When the neuron's membrane potential has already reached its peak before the update, the IMP should be updated using Equation (S.15) to reflect the cumulative effect of incoming inputs. If the neuron's membrane potential has not reached its peak before the update, the IMP should be updated using Equation (S.16) or (S.17) according to the magnitude of $t_{\text{n0}}$ and $t_{\text{n}} + \tau_{\text{m}}l_0 / 2\lambda$.

During continuous synaptic integration, the conditions that trigger the action potential also need to be considered. In the digital neuron, the excitability of the neuron should be considered during three periods (quiescence, refractory and subnormal period) with relatively long durations. During the quiescence period, the action potential will be triggered when the IMP satisfies the following condition

$$\max(V_{\text{m1}}(t))\ge V_{\text{th}}-V_{\text{rest}}, \quad \text{(S.18)}$$

where $V_{\text{th}}$ is the threshold of membrane potential and $V_{\text{rest}}$ is resting membrane potential. $V_{\text{th}}$ is typically 15 to 30 mV higher than $V_{\text{rest}}$. During the refractory period, the neuron will not trigger a new action potential regardless of the stimulus intensity. This period typically lasts for 0.5 to 1 ms. During the subnormal period, the neuron requires a stronger stimulus to reach the threshold potential since the membrane potential is below the resting potential. This period typically lasts for about 10 ms depending on the different neurons.

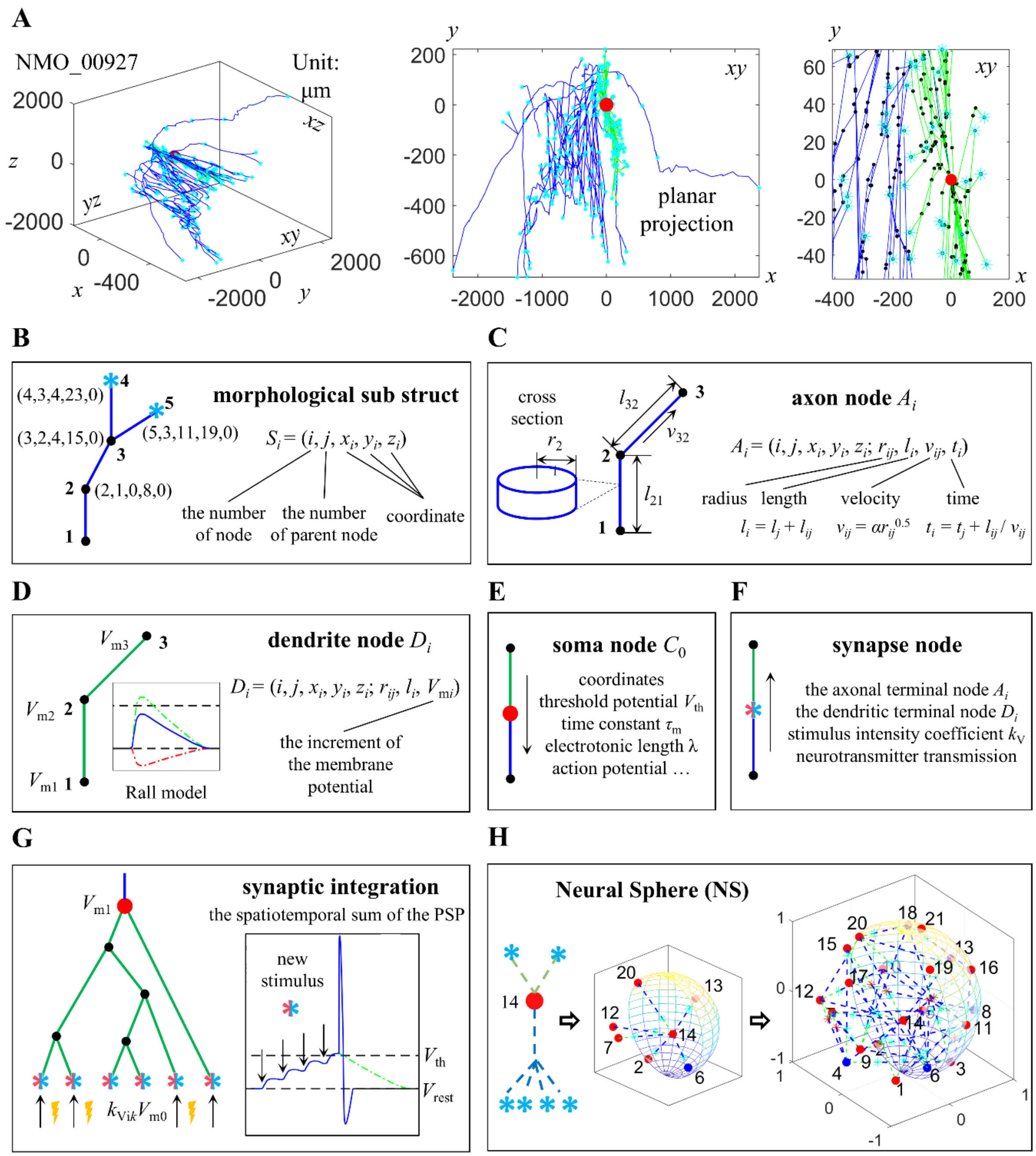

**Fig. S1.** The supplementary schematic diagram of the NS based on Fig. 1. (B-G) complement the composition of digital neuron. (A) Example of NMO_00927 excitatory neuron plotted through digital neuron. Morphological data for this neuron are obtained from Digitally Reconstructed Neurons and Glia(*17*). The blue lines represent the axon. The green lines represent the dendrites. The cyan asterisks indicate the synapse. The red dot indicates the soma. (left) Three-dimensional view of the NMO_00927 neuron. (middle) Projection of the NMO_00927 neuron on the *xy*-plane. (right) Local magnification near the soma. The black dots represent the nodes. (B) The schematic diagram of the morphological sub struct. (C) The schematic diagram of the axon node $A_i$. (D) The schematic diagram of the dendrite node $D_i$. It uses the Rall model(*27*) to describe the membrane potential on the dendrites. Green dotted line, red dotted line and solid blue line represent excitatory, inhibitory and sum postsynaptic potential (PSP), respectively. (E) The schematic diagram of the soma node $C_0$. The arrow indicates the direction of conduction. (F) The schematic diagram of the synapse node. (G) The schematic diagram of the synaptic integration. During synaptic continuously receives the stimulation, neuron determines whether to trigger an action potential by monitoring the spatiotemporal sum of the PSP. (H) The composition of the NS. It is constructed from projections and connections of digital neuron in the unit sphere. An example of the connection of neuron 14 in the sphere is presented here.

(2) Optimization of sphere distribution for NE

Determining the spatial distribution of neurons in the NE is crucial to achieve energy efficient communication. The three-dimensional distribution of densely connected small-scale NEs will be analyzed by mathematical optimization. Based on basic physical principles and biological phenomena, the arrangement of neurons can be parameterized into the following four criteria.

Firstly, neurons must not overlap with one another and require sufficient space to fulfill their normal metabolic requirements. Therefore, there is a minimum distance $l_{\min}$ between somas

$$(x_i - x_j)^2 + (y_i - y_j)^2 + (z_i - z_j)^2 \geq l_{\min}{}^2,\ 1 \leq i, j \leq n, i \neq j\ , \tag{S.19}$$

where $x_i$, $y_i$, $z_i$ are the coordinates of the soma of neuron $i$ ($1 \leq i \leq n$). The axons (or dendrites) are disjoint and at least $d_{\min}$ away from each other

$$(x'_{ip} - x'_{jq})^2 + (y'_{ip} - y'_{jq})^2 + (z'_{ip} - z'_{jq})^2 \geq d_{\min}{}^2,\ 1 \leq i, j \leq n, i \neq j\ , \tag{S.20}$$

where $x_{ip}$, $y_{ip}$, $z_{ip}$ are the coordinates of the point $p$ on neuron $i$.

Secondly, due to the limited space of the brain, the NE cannot be spread out indefinitely. Therefore, the NE is confined to a limited space

$$x_i^2 + y_i^2 + z_i^2 \leq R^2, \tag{S.21}$$

where $R$ is the radius of the restricted space. To make the optimization case unique, the centroid of the NE is translated to the origin

$$\sum x_i = 0, \sum y_i = 0, \sum z_i = 0\ . \tag{S.22}$$

Thirdly, the topological connection of NE should remain unchanged during optimization. The probability of connections between neurons should follow the observed data. The probability of connecting adjacent neurons in the hippocampus and other brain regions used for memory analysis is about 30% to 50%(*24*). Depending on the local small-scale NE, the probability of connection is about 5% to 80%, which is higher than the total probability of connection for the whole cortex.

Fourthly, to ensure lower energy consumption, the sum of the length of axons (or dendrites) should be as short as possible. To ensure faster communication, the total path between neurons (where the common axons (or dendrites) of different connections are counted repeatedly) should be as short as possible. However, in the brain, energy consumption and communication are considered simultaneously. The connections from a single neuron to multiple sites generally share a path first and bifurcate at the tail to form a tree of axon and ensure low metabolic energy consumption and fast communication. Here, the connection that guarantees energy efficiency is considered first. Suppose that the axon of neuron $i$ is linked to the dendrites of $x$ neurons. It will form a set of endpoints

$$T_i = \{N_i, N_{j1}, N_{j2}, \cdots N_{jx}\}, \tag{S.23}$$

where $N_i$ represents neuron $i$ as the root node of the tree of axon, $N_j$ ($j = j_1, j_2, \ldots, j_x$) represents the leaf nodes of the tree of axon. When the coordinates ($x_i$, $y_i$, $z_i$) of the root and leaf nodes are known, the coordinates of the intermediate nodes of the binary tree of axon should be optimized to find the shortest path. Suppose that the set of intermediate nodes is

$$S_i = \{S_{i1}, S_{i2}, \cdots, S_{ik}\}\ . \tag{S.24}$$

Then the set of nodes of the tree of axon is

$$V = T \bigcup S, \tag{S.25}$$

The set of edges of the tree is

$$E = \{(p, q) \mid p, q \in V\}\ . \tag{S.26}$$

The length of each edge is

$$L_{pq} = \sqrt{(x_p - x_q)^2 + (y_p - y_q)^2 + (z_p - z_q)^2} \ . \quad \text{(S.27)}$$

The objective function is to minimize the sum of the length of the edge

$$\min_{S(k)} L_{\text{total}} = \min_{S(k)} \sum_{(p,q)\in E} \sqrt{(x_p - x_q)^2 + (y_p - y_q)^2 + (z_p - z_q)^2} \ . \quad \text{(S.28)}$$

An important property of the optimal path is that the sum of the three edge vectors at each intermediate node is zero. This constraint can be written as

$$\sum_{S_h \in S, X_{hl} \in V} \frac{S_h - X_{hl}}{\| S_h - X_{hl} \|} = 0 , \quad \text{(S.29)}$$

where $X_{hl}$ is the node directly connected to $S_h$. Then the optimization equation can be expressed as

$$\begin{aligned} &\min_{S(k)} L_{\text{total}} = \min_{S(k)} \sum_{(p,q)\in E} \sqrt{(x_p - x_q)^2 + (y_p - y_q)^2 + (z_p - z_q)^2} , \\ &\text{subject to} \sum_{S_h \in S, X_l \in V} \frac{S_h - X_{hl}}{\| S_h - X_{hl} \|} = 0, \end{aligned} \quad \text{(S.30)}$$

According to the solution of the optimization equation (S.30), the shortest path for the axon of neuron $i$ to connect to the dendrites of $x$ neurons is defined as $L_{\min}(i, T_i)$.

Based on the above criterion, Spatial distribution of neurons in the NE of $n$ neurons with minimum energy consumption can be given by optimization. The objective function is to minimize the sum of the connection paths of all neurons. This optimization process is an NP-hard problem with high computational complexity. Using fast or conventional algorithms generally only converges to suboptimal solutions. When the topological connection is determined, the optimization equation is

$$\begin{aligned} &\min \sum_i L_{\min}(i, T_i) \\ &\text{subject to } (x_i - x_j)^2 + (y_i - y_j)^2 + (z_i - z_j)^2 \ge l_{\min}{}^2, \ 1 \le i, j \le n, i \ne j, \\ &\qquad (x'_{ip} - x'_{jq})^2 + (y'_{ip} - y'_{jq})^2 + (z'_{ip} - z'_{jq})^2 \ge d_{\min}{}^2, \ 1 \le i, j \le n, i \ne j, \\ &\qquad \sum x_i = 0, \sum y_i = 0, \sum z_i = 0, \\ &\qquad x_i^2 + y_i^2 + z_i^2 \le R^2. \end{aligned} \quad \text{(S.31)}$$

In the case of full connection ($p$ = 100%), optimization is performed by Equation (S.31). The optimal spatial distribution of neurons will constitute a sphere structure as shown in Fig. S2. The soma nodes are evenly distributed on the surface of the sphere following the Fibonacci grid(*28*). It is not only more consistent with the actual situation, but also beneficial to numerical calculation. In other topologically connected NEs, sphere structures are at least a suboptimal case. Therefore, when NS is applied to small-scale NEs with high connection probability, the single-layer sphere structure can ensure high communication efficiency while taking into account the structural order.

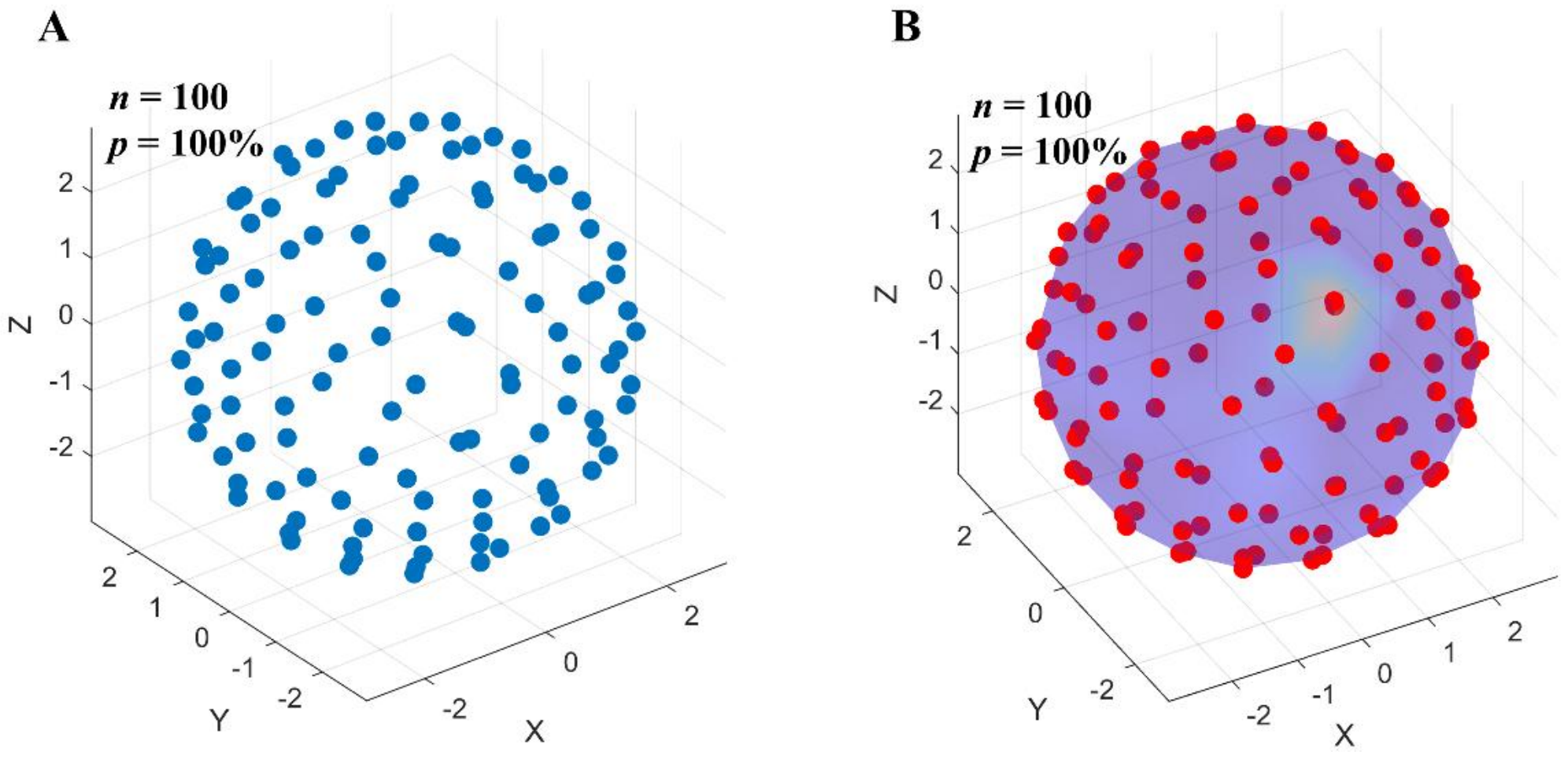

**Fig. S2.** Optimal distribution of spatial coordinates of neurons with $n = 100$, $p = 100\%$. (A) Optimal spatial distribution of neuronal somas. (B) Sphere structure of the coordinate distribution for the NE. The detailed visualization of the optimization process is presented in Movie S1.

(3) Connection structure for NS

Based on the digital neurons and optimization conclusions, connecting multiple digital neurons into a structured spherical network can maximally simulate the energy efficient electrophysiological activity of NE. This model can be named figuratively as the neural sphere (NS). The primary structure of the NS is a unit sphere. The soma nodes of each digital neuron are distributed on the surface of the sphere, while the connections between neurons (axons → synapses → dendrites) are distributed within the sphere as shown in Fig. S1H.

The NS can be used to simulate real neural networks, such as a small ensemble of neurons in the cerebral cortex. In the NS, the sphere surface is equivalent to an unfolded cerebral cortex. Using topologically equivalent rules, the positions and connections of neurons are mapped onto the surface and inside of the sphere.

In addition, NS can also be used to design specialized ideal networks to study the principles of bio-intelligence. In a uniform ideal NS, the soma nodes are evenly distributed on the surface of the sphere following the Fibonacci grid(*28*). Then the coordinates of $n$ neurons on the sphere can be expressed as

$$\begin{cases} z_i = \dfrac{2i-1}{n} - 1, \\ x_i = \sqrt{1-z_i^2}\cos(2\pi i\phi), \\ y_i = \sqrt{1-z_i^2}\sin(2\pi i\phi), \end{cases} \tag{S.32}$$

where $i = 1, 2, \ldots, n$ and constant $\phi = (5^{1/2}-1)/2$. By setting the parameters of the digital neurons and the neuron connection matrix, NSs with different functions and the characteristics of electrical activity can be obtained.

Implementation of numerical experiments

Based on NS, the numerical simulation of neural network action potential activity can be implemented by C++. This section will briefly describe the implementation process of the simulation program.

The program inputs include a basic parameter table, initial conditions, digital neuron files, and a neuron synaptic connection file. The basic parameter table is used to set global parameters such as the number of neurons, the maximum simulation time, and the time step. The initial conditions specify the excitation state of each neuron at the beginning, including states such as resting, pre-action potential, during action potential, post-action potential and so on. Each neuron in the network has an independent digital neuron csv file.

After receiving input, the state of each node in every neuron will be simulated at each time step. The simulation process proceeds as follows: update the dendritic input (update dendrite nodes), update the membrane potential (synaptic integration), and update the axonal output (output action potentials). After reaching the maximum simulation time, the program will output the membrane potential of each neuron over time.

In all numerical experiments, the following morphological parameters of all neurons are assigned the same values, including soma radius, dendrite radius, axon radius, average dendrite length, and average axon length. These parameters are obtained through statistical analysis of the morphological data of pyramidal neurons in the monkey cerebral cortex from the digitally reconstructed neurons and glia database, neuromorpho.org(*17*). The reduced Traub-Miles model(*11*), a variant of Hodgkin-Huxley action potential model(*6-10*) for pyramidal cells, is used to simulate the action potential of neurons. In some of the numerical experiments, randomly connected neural networks are employed. The random parameters in the randomly connected network include synaptic connection strength, neuron connection probability $p$, etc. Unless otherwise specified, the initial condition is that the first neuron triggers an action potential at 0.00 ms, the maximum simulation time is 1000 ms, and the time step is 0.01 ms. The detailed parameter settings are presented in Supplementary Text.

Correlation analysis of membrane potential time series

In order to determine the correlation between the membrane potential time series of two neurons, it is necessary to design a reasonable correlation analysis method. Suppose that the membrane potential time series of two neurons $i$ and $j$ with the same time length are $V_i(t)$ and $V_j(t)$, respectively. Due to the different initial conditions, even the same two membrane potential time series will have a time difference. Therefore, it is necessary to translate one of the time series by time $t_0$, and then calculate the Pearson correlation coefficient $\rho$ of the overlap part of the two time series.

The correlation coefficient $\rho_{\mathrm{m}}$ between $V_i(t)$ and $V_j(t)$ can be expressed as

$$\rho_{\mathrm{m}}(V_i(t), V_j(t)) = \max_{t_0}\{\rho(V_i(t-t_0), V_j(t))\},\ 0 \le t_0 \le T\ , \tag{S.33}$$

where $T$ is the period of the membrane potential time series.

Hysteresis complex iteration theory of simplified NS dynamical system

It is necessary to characterize the dynamical principles of NS to drive different potential activities. Depending on the initial conditions, NS can generate different types of potential activity to simulate the electrophysiological activity of the NE. Since the full initial condition simulation through the original NS requires unacceptable computing power and time, a simplified NS dynamical system can be constructed to restore the main biological links of electrophysiological activities in NE. The simplified NS dynamical system consists of the simplified action potential, the simplified synaptic connection, and the hysteretic complex iterative process.

The simplified action potentials can be reduced to the basic feature of all-or-nothing by a step function $\varepsilon$ and a Dirac delta function $\delta$, as shown in Fig. S3. The simplified action potential function of a neuron can be expressed as

$$U(\tau) = \varepsilon(\tau_{\mathrm{r}} + t - \tau)\delta(\tau - t), \tag{S.34}$$

where $U$ is the equivalent state of the membrane potential of the neuron, $\tau_{\mathrm{r}}$ is absolute refractory period and $t$ is the moment when the action potential is triggered.

The simplified synaptic connections can be represented by the connection matrix $\boldsymbol{K}_{\mathrm{V}}$

$$\boldsymbol{K}_{\mathrm{V}} = \begin{bmatrix} \boldsymbol{K}_{\mathrm{V},1} \\ \boldsymbol{K}_{\mathrm{V},2} \\ \vdots \\ \boldsymbol{K}_{\mathrm{V},n} \end{bmatrix} = \begin{bmatrix} K_{11} & K_{12} & \cdots & K_{1n} \\ K_{21} & K_{22} & \cdots & K_{2n} \\ \vdots & \vdots & \ddots & \vdots \\ K_{n1} & K_{n2} & \cdots & K_{nn} \end{bmatrix}, \tag{S.35}$$

where the real part of $K_{ij}$ represents the sum of connection strengths of all synapses from the axon of neuron $i$ to the dendrite of neuron $j$, and the imaginary part of $K_{ij}$ represents the hysteretic time of signal transmission from the axon to the dendrite. In particular, $K_{ij}$ is 0 when there is no connection.

According to the above simplification, the simplified NS dynamical equation can be constructed to essentially describe the dynamic process of continuous hysteretic iterative interaction among neurons. The state variable of each neuron can be described by a complex number

$$z = V(\cos\tau + i\sin\tau), \tag{S.36}$$

where $V$ is the equivalent potential state of the neuron and $\tau$ is time. Since $\tau$ is a small quantity compared to $V$, the following approximation can be made when plotting the phase portrait

$$z \approx V + i\tau. \tag{S.37}$$

Therefore, the simplified NS dynamical equation can be expressed as

$$z_{k+1,i}(\tau) = (a_i - z_{k,i}(\tau))g(K_{ij}z_{k,j}(\tau) - \Phi_i)\varepsilon(\tau_{\mathrm{r}} + kt_{\min} - \tau)\delta(\tau - kt_{\min}), \tag{S.38}$$

where $i$ is the index of the neuron and $g$ is the sigmoid function. This tensor expression consists of a control term $(a_i - z_{k,i})\, g(K_{ij}z_{k,j} - \Phi_i)$ and an action potential term $\varepsilon(\tau_{\mathrm{r}} + kt_{\min} - \tau)\, \delta(\tau - kt_{\min})$, where $k$ is the iteration step and $t_{\min}$ is the time resolution. The real part of $a_i$, Re $(a_i)$, is used to regulate whether the neuron triggers an action potential. Depending on the change of Re($z$), Re $(a_i)$ made the control term is close to 1 or 0. The imaginary part of $a_i$, Im $(a_i)$, is used to set the hysteretic time of synaptic integration for neuron $i$. The real number $\Phi_i$ is the threshold of neuron $i$. Each complex iteration of the dynamical system is equivalent to the synaptic integration of each neuron for several hysteresis inputs from the remaining connected neurons. The dynamics of the simplified NS dynamical system is regulated by three groups of control parameters, including $a_i$, $K_{ij}$ and $\Phi_i$. If only the simplest interaction of neurons in NS is considered (that is, each neuron is only subjected to the hysteresis effect of the connected neurons), the hysteresis dynamical equation of NS can be further simplified as

$$z_{k+1,i}(\tau) = (a_i' - z_{k,i}(\tau))(K_{ij}' z_{k,j}(\tau) - \Phi_i'), \tag{S.39}$$

where $a_i'$, $K_{ij}'$ and $\Phi_i'$ are the equivalent control parameters of $a_i$, $K_{ij}$ and $\Phi_i$ after affine transformation, respectively.

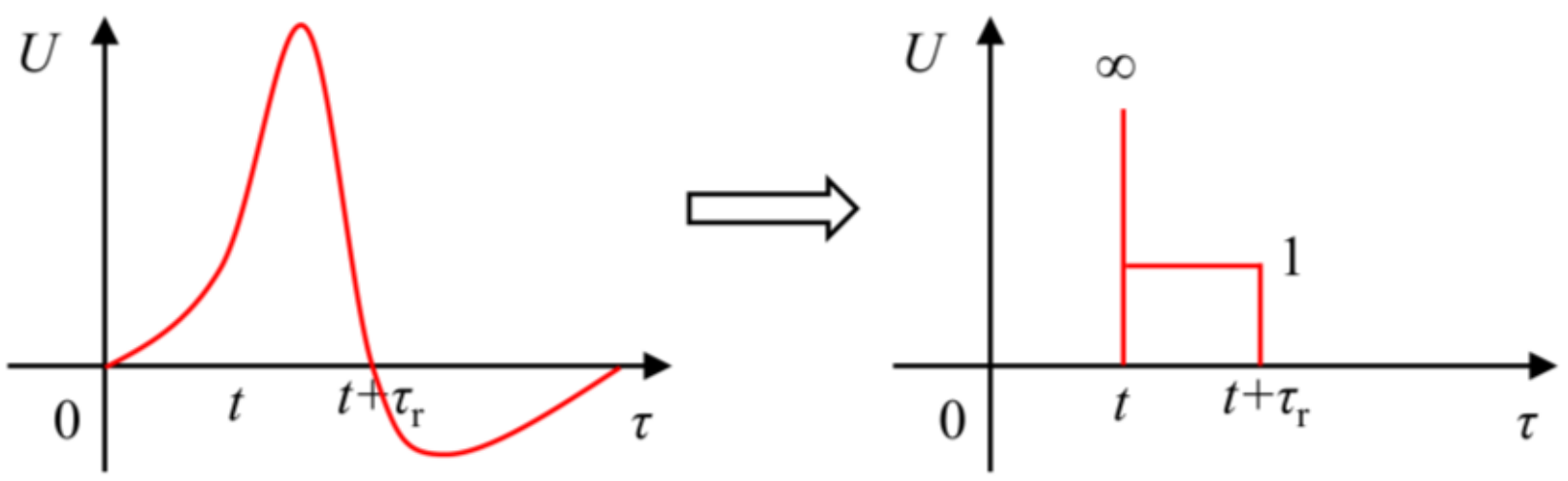

**Fig. S3.** Simplified schematic representation of the action potential for the neuron.

Quantification of information in the HNCA and the artificial systems

Using the principles of information theory, the data storage capacity of the memory NS and the artificial system are quantified. Detailed estimations are presented in Supplementary Text.

In 1948, Shannon used the concept of entropy to quantify information and named it self-information(*32*). The self-information $I$ is represented as

$$I = -\sum_i p_i \log_2 p_i ,\ i = 1, 2, \cdots, q , \tag{S.40}$$

where $p_i$ is the probability that event $i$ occurs and $q$ is the total number of events.

In current computing architectures, manual computing systems for large-scale applications are all binary encoded. Each device has only two states. In this case, the self-information of these artificial systems can be represented as

$$I = -\sum_i (p_i \log_2 p_i + (1 - p_i)\log_2(1 - p_i)),\ i = 1, 2, \cdots, n , \tag{S.41}$$

where $p_i$ is the probability that the device $i$ is in one of the binary states. The value of $I$ is from 0 to $n$ bit.

In the high-performance neuromorphic computing architecture of brain (HNCA), the number of events in the self-information corresponds to the number of periodic potential waveforms. And each periodic potential waveform corresponds to an initial condition group. If the state of a neuron at a certain time is intercepted as an initial condition, the number of initial states of a neuron consists of four terms

$$\begin{gathered} S_{\text{neuron}} = S_{\text{Q}} + S_{\text{I}} + S_{\text{R}} + S_{\text{S}} , \\ S_{\text{Q}} = 1,\ S_{\text{I}} = \frac{V_{\text{th}} - V_{\text{rest}}}{\max(V_{\text{m0}})} \frac{\tau}{t_{\min}},\ S_{\text{R}} = \frac{t_{\text{R}}}{t_{\min}},\ S_{\text{S}} = \frac{t_{\text{S}}}{t_{\min}} , \end{gathered} \tag{S.42}$$

where $S_{\text{Q}}$, $S_{\text{I}}$, $S_{\text{R}}$ and $S_{\text{S}}$ are the number of states of the neuron during quiescence, incubation, refractory, and subnormal periods, respectively. $V_{\text{th}}$ is the threshold of membrane potential and $V_{\text{rest}}$ is resting membrane potential. $\max(V_{\text{m0}})$ is the peak of the membrane potential generated by one vesicle. $\tau$ is the time when the membrane potential reaches its peak from 0. $t_{\text{R}}$ and $t_{\text{S}}$ are the durations of the refractory and subnormal periods, respectively. $t_{\min}$ is the minimum temporal resolution of the neuron. If the minimum temporal resolution of the neurons can reach the physical limit, $t_{\min}$ will take Planck time, $5.39116(13)\times10^{-44}$ s. Then the total number of initial conditions of the memory NS consisting of $n$ neurons is

$$S_{\text{ensemble}} = (S_{\text{neuron}})^{s} , \tag{S.43}$$

where $s$ ($\geq$ n) represents the number of external synapses that trigger the action potential activity of the ensemble. The number of initial conditions corresponding to different periodic potential waveforms is also different. The average number of initial conditions of the periodic potential waveform can be expressed as

$$S_{\text{T}} = \frac{\sum_j T_j / t_{\min}}{S_{\text{ensemble}}},\ j = 1, 2, \cdots, S_{\text{ensemble}} , \tag{S.44}$$

where $T_j$ is the period of action potential activity at initial condition $j$. Besides, the synaptic plasticity in the HNCA will multiply the number of initial conditions. The total number of states of external synapses can be defined as

$$S_{\text{synapse}} = (n_{\text{v}} + 1)^{s} , \tag{S.45}$$

where $n_v$ represents the total number of states for synaptic release vesicles. "+1" represents a state in which no vesicles are released. Then the number of periodic potential waveforms contained by the HNCA can be estimated as

$$S_{total} = \frac{S_{ensemble} \bullet S_{synapse}}{S_T}, \quad (S.46)$$

In a set of NS, the number of states of the HNCA can be estimated as

$$S_{total} = \prod_{j=1}^{m} \frac{S_{ensemble}(j) \bullet S_{synapse}(j)}{S_T(j)}, \quad (S.47)$$

where $m$ is the number of sub NS. The $S(j)$ with different subscripts represent the corresponding number of states of the sub NS $j$.

As for the HNCA of the fractal of $m' \times m$ (Fig. 4), the number of states of the HNCA can be estimated as the product of three terms

$$S_{total} = (\prod_{j=1}^{m'} S_{total}(j))(S_{synapse} - \prod_{j=1}^{m'} S_{synapse}(j))(\prod_{j=1}^{m'} \frac{S_T(j)}{t_{min}}), \quad (S.48)$$

where the first term represents the sum of the number of sub memory NS states, the second term represents the synaptic plasticity of the connections among sub memory NSs, and the third term represents the number of states of the periodic potential waveform of the sub memory NS. Therefore, the self-information of the memory NS of $n$ neurons is

$$I' = -\sum_{i} (p_i' \log_2 p_i'), \ i = 1, 2, \cdots, S_{total}, \quad (S.49)$$

where $p_i'$ is the probability of generating waveform $i$.

For fractal structures with more layers (>2), Equation (S.43) can be rewritten as

$$S_{ensemble}{}^{(p+1)} = (S_{total}{}^{(p)})^{s^{(p+2)}}, \quad (S.50)$$

where $S_{ensemble}{}^{(p+1)}$ is the number of states of the ensemble of fractal at level $p+1$. $S_{total}{}^{(p)}$ is the total number of states of fractal at level $p$. $s^{(p+2)}$ is the number of synaptic connections between sub ensembles in the fractal at level $p+2$.

Besides, according to the analysis of the NS with periodic potential activity, the number (or self-information) of initial conditions will be greater than the number (or self-information) of periodic potential waveforms in NS. The number of distinct initial condition matrices as shown in Fig. 2E is equivalent to the number of states of the initial condition. Its limitations include the following four aspects. Firstly, the time at which the next action potential triggers will be after the last one in the same neuron, i.e., $t_{ji} > t_{ki}, j > k$. Secondly, the time difference between two adjacent triggered action potentials is greater than the refractory period in the same neuron, i.e., $t_{ji} > t_R$. Thirdly, the final $t_{ji}$ is supplemented with "/" when the action potential is fired less than $q$ times. Fourthly, the final non-"/" $t_{ji}$ should be less than the period of the periodic potential waveform of NS, i.e., $t_{ji} < T$. Based on the above limitations, the number of the initial condition matrices is

$$S_{matrix} = \left(\sum_{j=1}^{q} \frac{([\frac{T}{t_{min}}] - (j-1)[\frac{t_R}{t_{min}}])!}{j!([\frac{T}{t_{min}}] - (j-1)[\frac{t_R}{t_{min}}] - j)!}\right)^n, \quad (S.51)$$

where

$$q = [[\frac{T}{t_{min}}] \Big/ [\frac{t_R}{t_{min}}] + 1], \quad (S.52)$$

and "[ ]" represents the rounding operation. Then the self-information of the initial condition is

$$I'' = -\sum_i (p_i'' \log_2 p_i''),\ i = 1, 2, \cdots, S_{\text{matrix}}. \tag{S.53}$$

Then the ratio of *I*'' and *I*' can measure the information compression rate or feature extraction amount of NS.

Evaluation of the computational power for HNCA

The computational power for floating-point operation of HNCA can be estimated by two equivalent methods. Detailed estimations are presented in Supplementary Text.

In NSs with random action potential activity, the action potential output of one neuron can be equivalent to several floating-point operations. The computational power of HNCA can be expressed as

$$\text{FLOPS}_{\text{HNCA}} = n \times \text{FLOP}_{\text{HNCA}} \times f, \tag{S.54}$$

where $\text{FLOP}_{\text{HNCA}}$ is the number of floating point operations equivalent to a single action potential. $n$ is the number of the neuron. $f$ is the frequency of the action potential. An action potential spike is the result of dendritic synaptic integration of that neuron. Suppose that the input to each synapse is equivalent to a floating point number. Then an action potential is equivalent to (the number of synapses -1) floating-point operations. in this equivalent procedure, the precision of one floating-point operation is $\log_2(\tau_m / t_{\min})$.

In NSs with periodic action potential activity, the total amount of memory operated per unit time can be equivalent to several floating-point operations. The computational power of HNCA can be expressed as

$$\text{FLOPS}_{\text{HNCA}} = \frac{I}{3\mu T}, \tag{S.55}$$

where $I$ is the self-information or storage capacity of the NS. "3" represents the occupancy that requires two inputs and one output for one floating-point operation. $\mu$ is the precision of a floating-point number. A double precision floating-point number has 64 bit of precision. $T$ is the period of action potential activity of the NS.

Physical limit of the power for human brain by HNCA

Based on Landauer's principle(*34, 35*), the power physical limit of human brain simulated by HNCA can be analyzed. In Landauer's principle, the energy required to erase a bit has a minimum value, which is called Landauer limit

$$E_{\min} = k_B T \ln 2, \tag{S.56}$$

where $k_B$ is the Boltzmann constant and $T$ is the temperature of system in Kelvin. Based on the estimation of the maximum computational power of the human brain $P_{\max}$ by HNCA, the theoretical minimum power of the human brain can be obtained by the Landauer limit

$$\begin{aligned} E_{\text{human brain}} &= k_B T \ln 2 \times P_{\max} \times 2\varepsilon_P \\ &= 1.38 \times 10^{-23} \times (273.15 + 37) \times \ln 2 \times 6.24 \times 10^{18} \times 2 \times 64 = 2.37\ \text{W}, \end{aligned} \tag{S.57}$$

where T takes the body temperature 37 °C, $2\varepsilon_P$ is the number of bits occupied by a double-precision floating-point operation.

Based on the existing experimental data, the energy composition of the human brain (15~20 W) can be divided into three parts(*36*), including computing and communication (≈ 3 W), maintaining body temperature (≈ 9 W), and energy consumption for other biological processes such as metabolism (≈ 3~8 W). Among them, the computing and communication power of the human brain is only 1.26 times that of the Landauer limit.

## Supplementary Text

The construction and parameter selection of NS

This section is to supplement "Neural sphere for simulating the brain" and "Implementation of numerical experiments" in the description of the NS.

The parameter settings for numerical simulations are presented in Table S1. All values are estimated from statistical analyses of experimental observations(*17*) of real neurons. Morphological data of some neurons are presented in Table S2. The parameters of the action potentials were derived from the RTM model(*11*) for the pyramidal cells as shown in Fig. S4.

| Type | Parameter | Description | Value |
|---|---|---|---|
| Simulation time parameters | $t_{max}$ | The maximum simulation time | 1000 ms |
| | $t_{min}$ | The time step | 0.01 ms |
| Initial condition | / | Neurons trigger action potential spectra sequentially | / |
| Neural ensemble parameters (the monkey cerebral cortex digital neuron database(*17*)) | $n$ | The number of neurons | / |
| | $s$ | The number of synapses | / |
| | $p$ | The probability of connection | 0-100% |
| | $\eta_i$ | The proportion of inhibitory neurons | 5%-15% |
| Parameters of pyramidal cells in monkey cerebral cortex(*17*) | $V_{th}$ | The threshold potential that triggers an action potential | -50 mV |
| | $V_{rest}$ | Resting potential | -70 mV |
| | $\tau_m$ | Membrane time constant | 20 ms |
| | $\lambda$ | Membrane length constant | 1000 μm |
| | $t_s$ | Synaptic response time | 0.5 ms |
| | $l_a$ | Average axon length based on database statistics | 700 μm |
| | $l_d$ | Average dendrite length based on database statistics | 300 μm |
| | $r_a$ | Average axon radius based on database statistics | 0.2 μm |
| | $r_d$ | Average dendrite radius based on database statistics | 0.55 μm |
| Action potential parameters | $t_r$ | Absolutely refractory period | 0.61 ms |
| | $t_s$ | Subnormal period | 11.73 ms |

**Table S1.** The parameter settings for numerical simulations based on NS.

| NMO index | Radius of dendrite (μm) | Length of dendrite (μm) | Radius of axon (μm) | Length of axon (μm) |
|---|---|---|---|---|
| 1 (01911) | 0.527328 | 264.965 | 0.37918 | 621.087 |
| 2 (35129) | 0.557319 | 301.176 | 0.254106 | 1256.24 |
| 3 (35130) | 0.427115 | 330.985 | 0.219363 | 473.237 |
| 4 (149530) | 0.528261 | 226.785 | 0.144715 | 491.611 |
| 5 (149531) | 0.728944 | 318.901 | 0.211443 | 804.431 |
| 6 (149532) | 0.700106 | 370.764 | 0.261524 | 1043.58 |
| 7 (149535) | 0.506562 | 255.191 | 0.18006 | 972.19 |
| 8 (149537) | 0.614571 | 310.844 | 0.256278 | 549.454 |
| 9 (149538) | 0.522 | 184.208 | 0.213849 | 414.821 |
| average | 0.568023 | 284.8688 | 0.235613 | 736.2946 |
| median | 0.528261 | 301.176 | 0.219363 | 621.087 |

**Table S2.** Morphological data of some of pyramidal cells in monkey cerebral cortex.

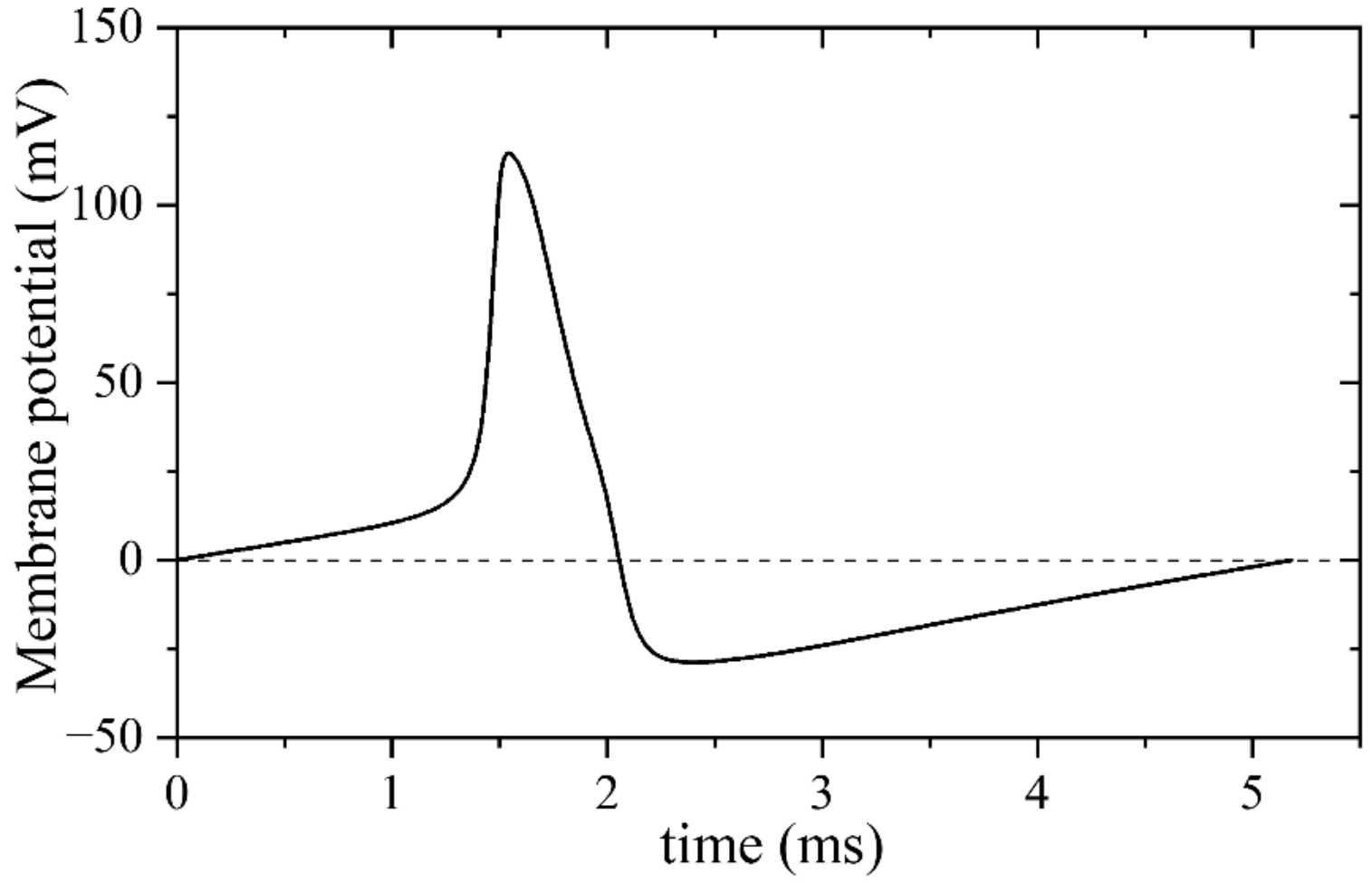

**Fig. S4.** Action potentials of pyramidal cells by RTM model(*11*).

Membrane potential activity data for NSs with different connection probabilities

Comprehensive membrane potential activity data in Fig. 2 are presented in this section. Figures S5-S8 correspond to the four subfigures of Fig. 2, respectively. The solid green line is the membrane potential of one of the neurons as a function of time. The black dashed line is the threshold potential, 20 mV. The blue solid line is the real-time frequency spectra of action potential as a function of time. The red solid line is the curve of the autocorrelation coefficient with time. The minimum period of the potential activity is at the first non-noise significant peak of the autocorrelation coefficient.

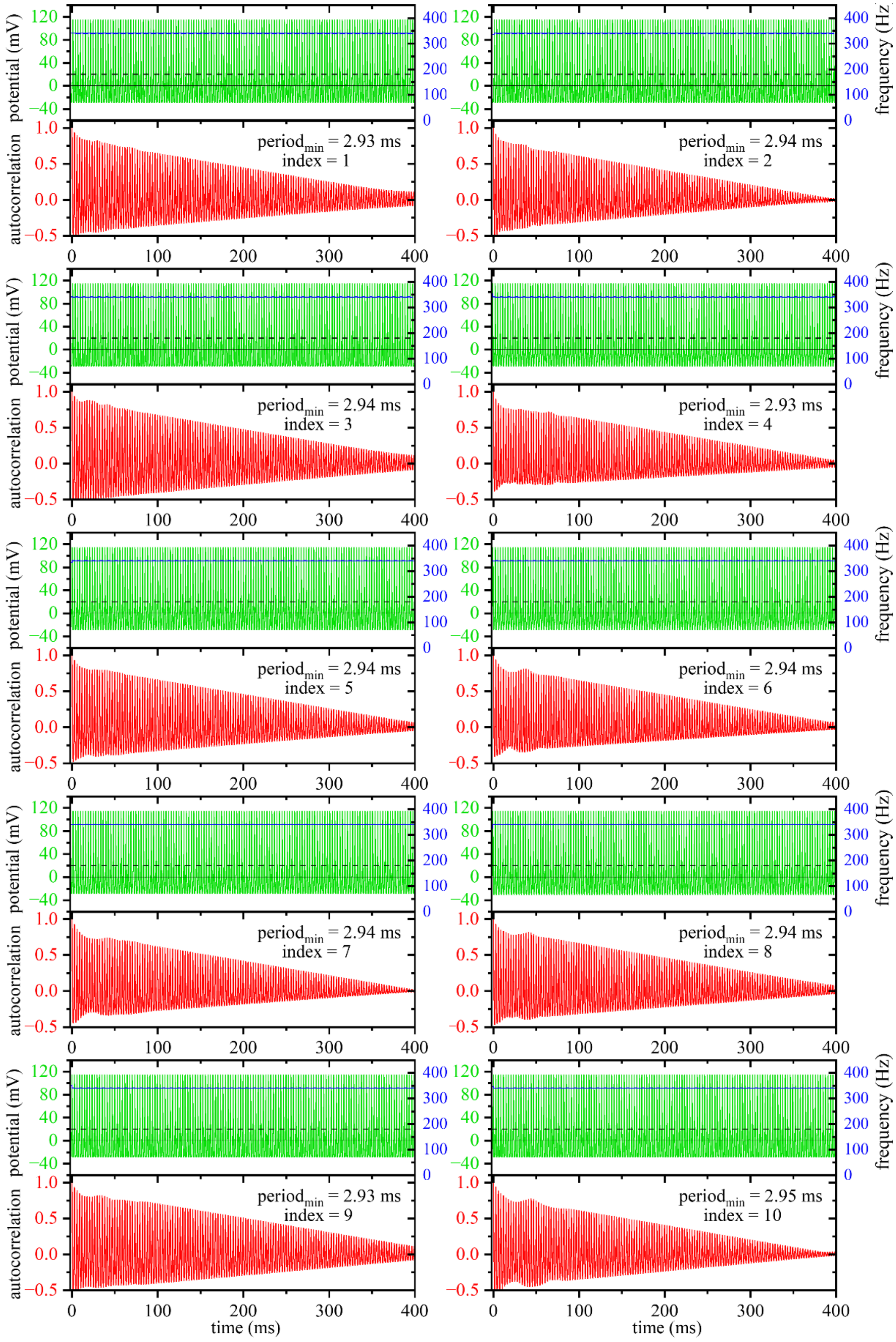

**Fig. S5.** Data for all neurons of the bidirectional fully connected NS.

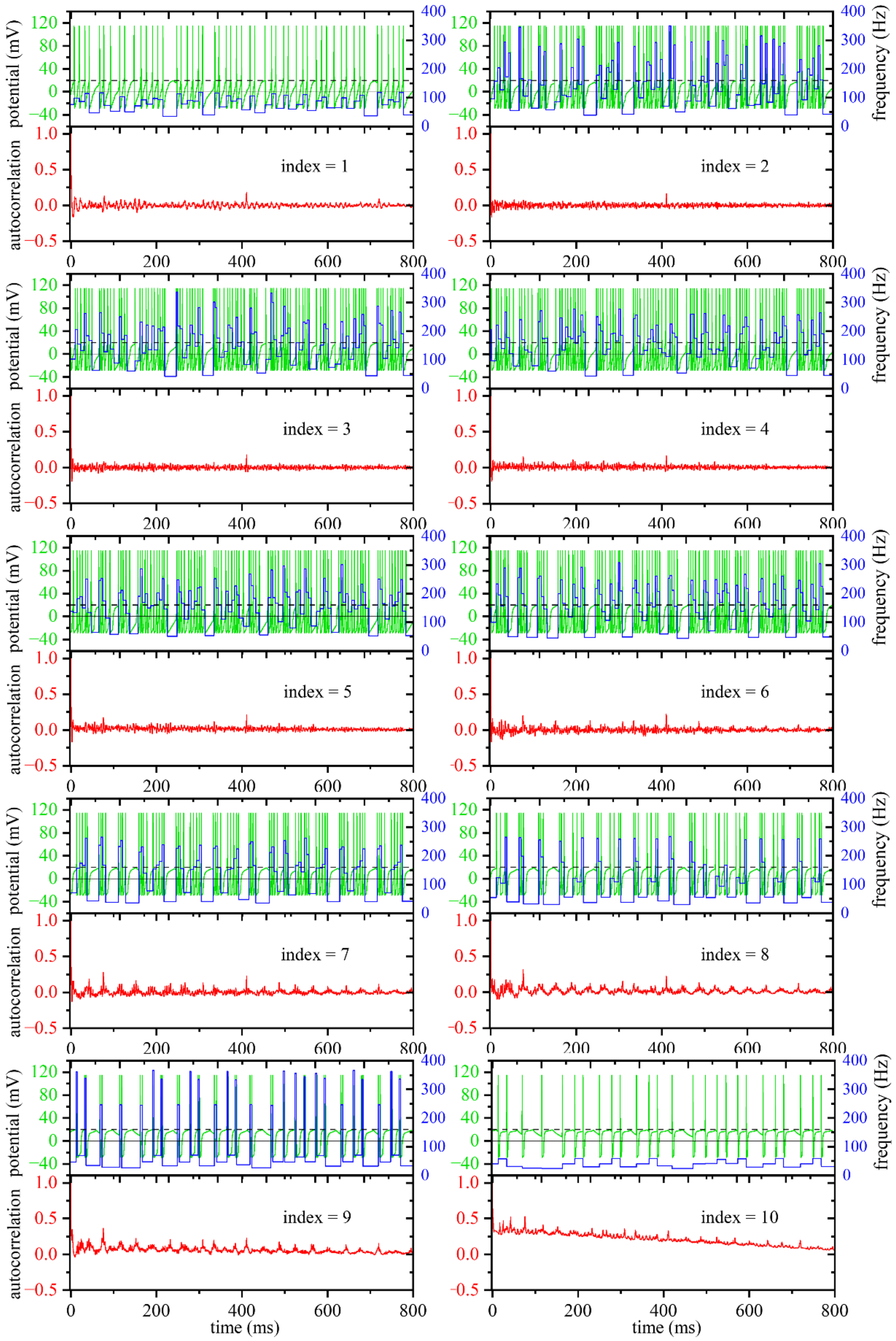

**Fig. S6.** Data for all neurons of the unidirectional fully connected NS.

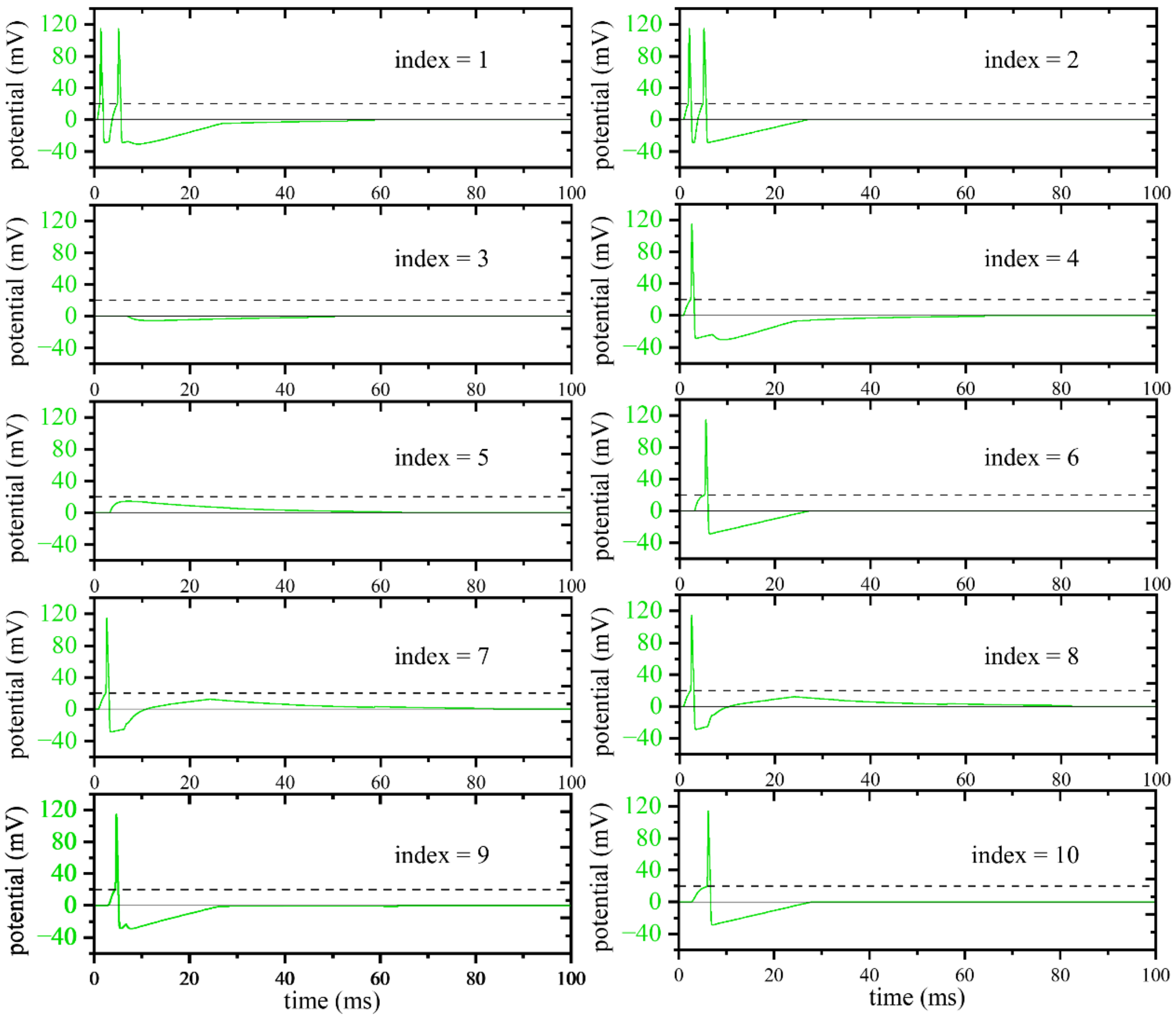

**Fig. S7.** Data for all neurons of the sparsely connected NS.

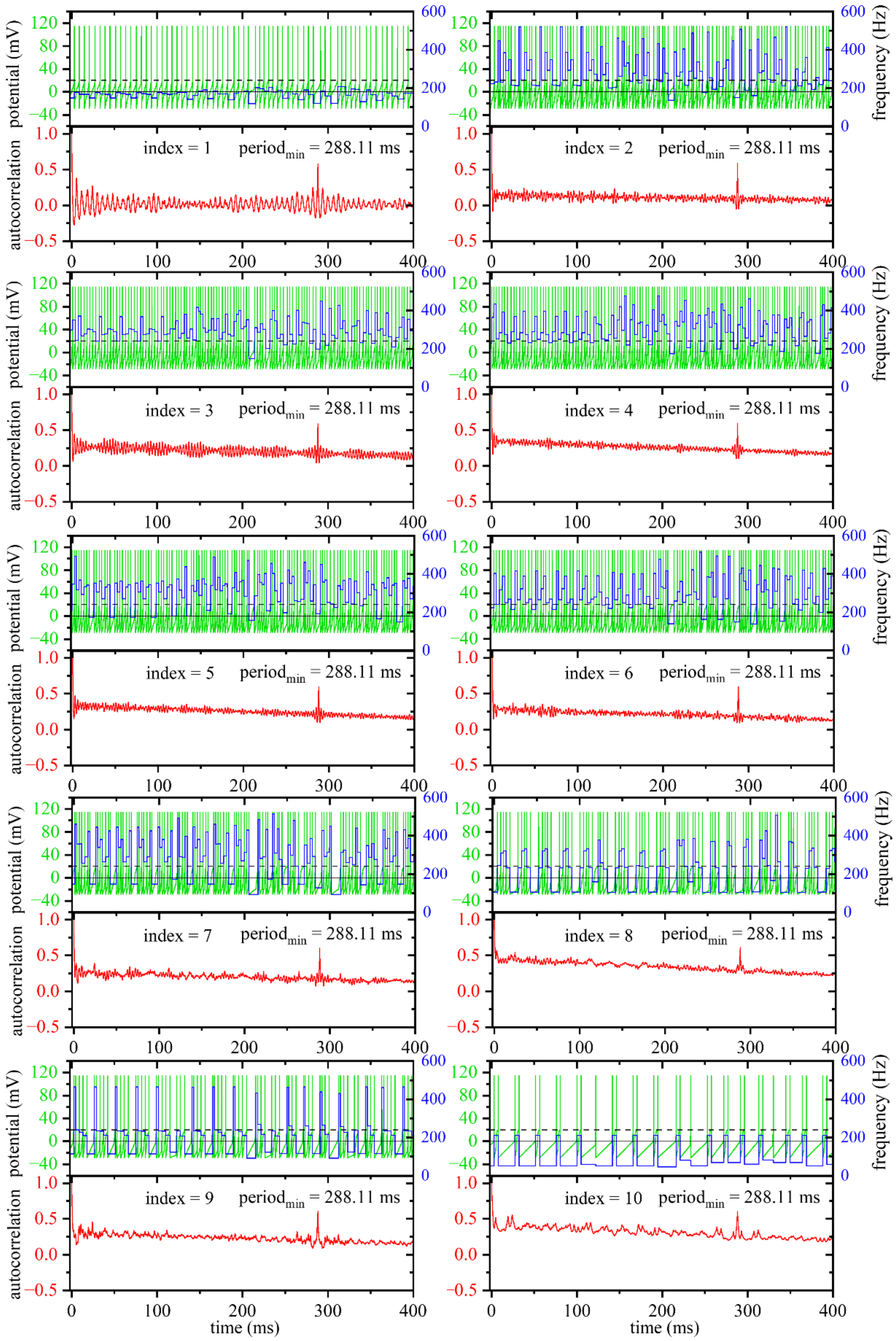

**Fig. S8.** Data for all neurons of the full positively feedback connected NS.

Membrane potential activity data for the full positively feedback connected NSs

Figures S9-S17 are the membrane potential activity data for the full positively feedback connected NS of nine different initial conditions to supplement Fig. 3F. When the initial condition is that the 10th neuron triggers an action potential at 0.00 ms, NS cannot generate continuous electrophysiological activities. Therefore, this situation is not recorded.

Figure S18 shows the autocorrelation coefficients of the potential activity of the above full positively feedback connected NSs. With the change of initial condition, the full positively feedback connected NSs have a finite number of different periods. Most of the periods are 288.11 ms.

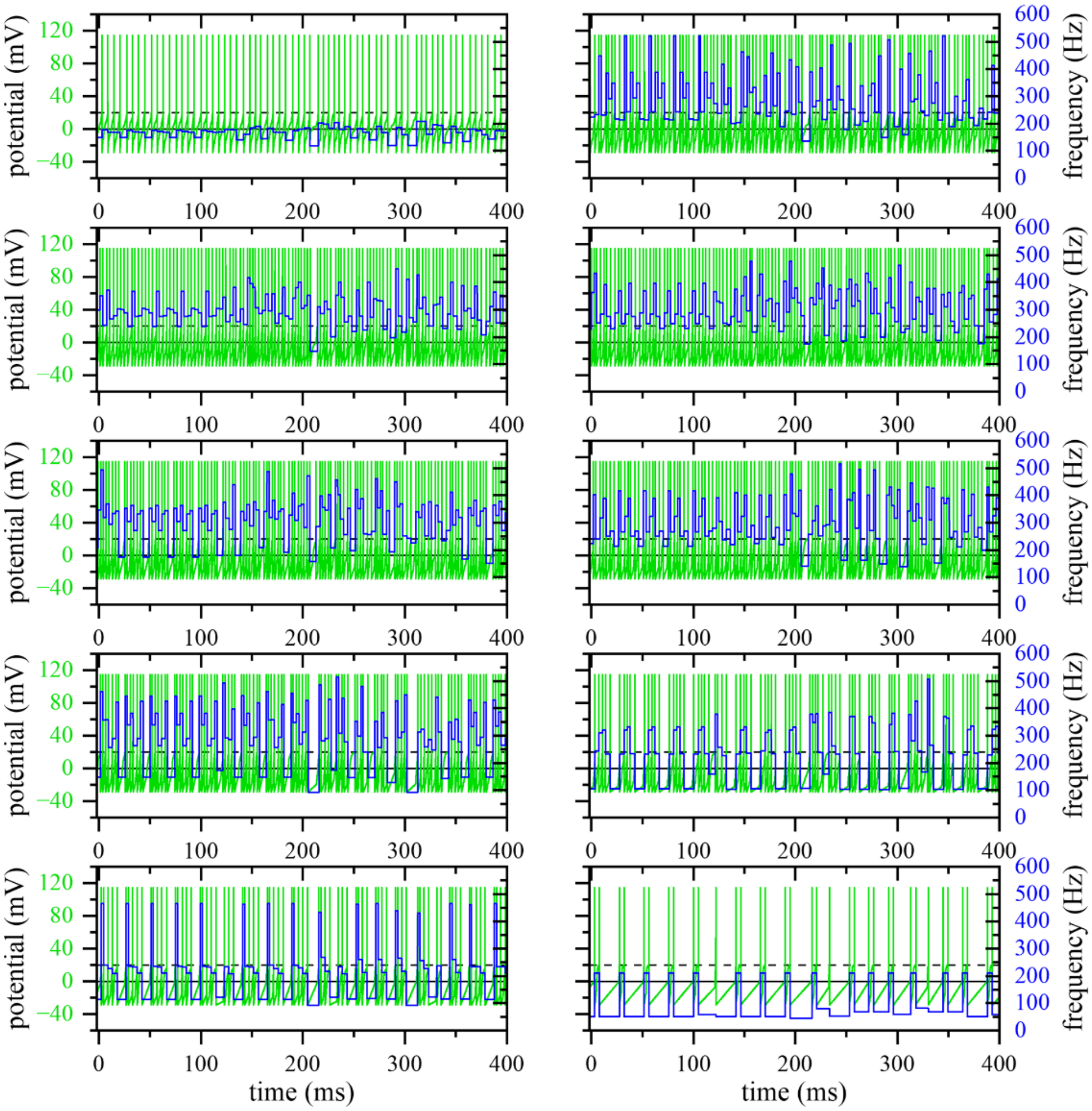

**Fig. S9.** Potential activity of the full positively feedback connected NS of 10 neurons. The initial condition is that the 1st neuron triggers an action potential at 0.00 ms.

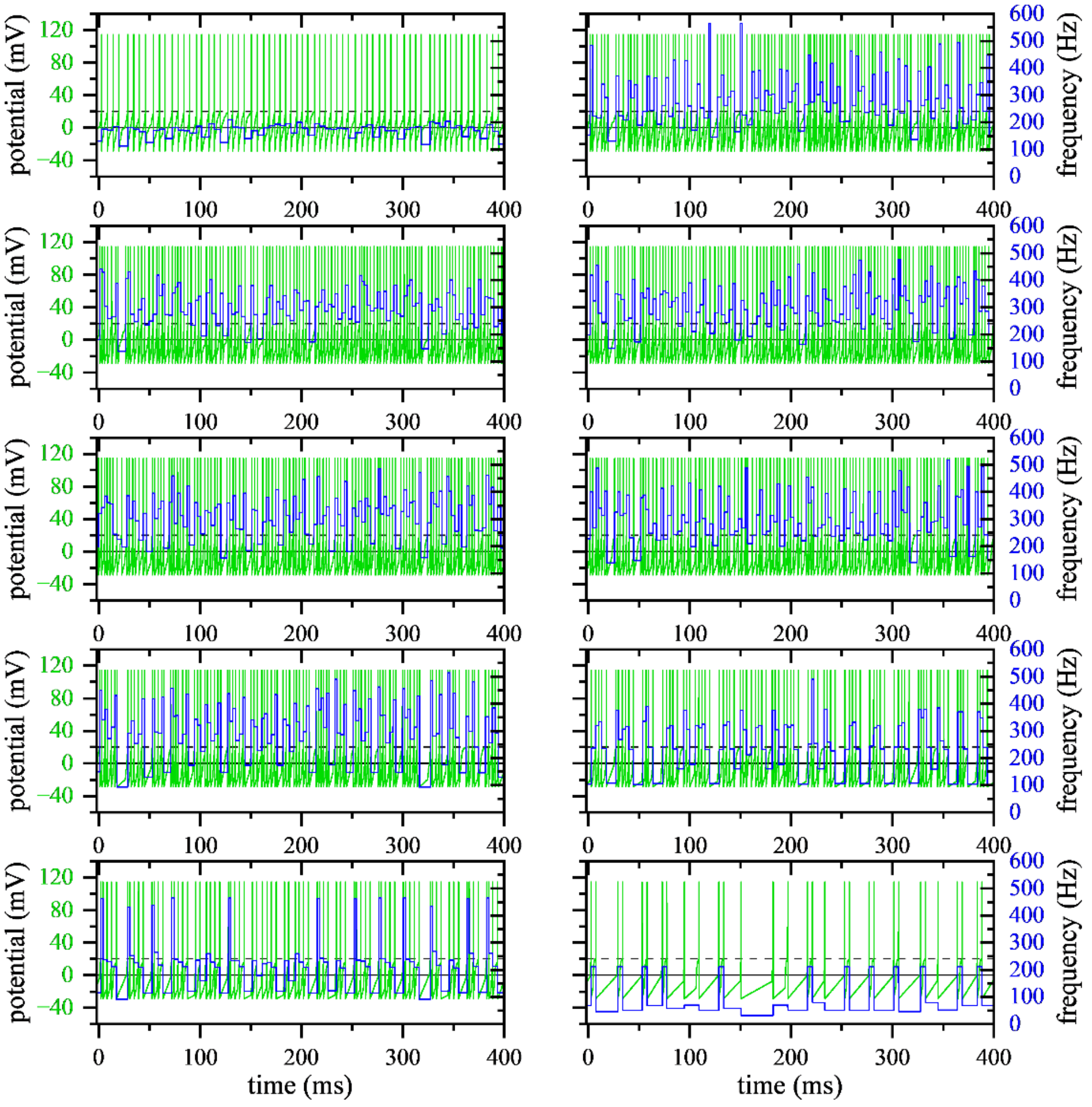

**Fig. S10.** Potential activity of the full positively feedback connected NS of 10 neurons. The initial condition is that the 2nd neuron triggers an action potential at 0.00 ms.

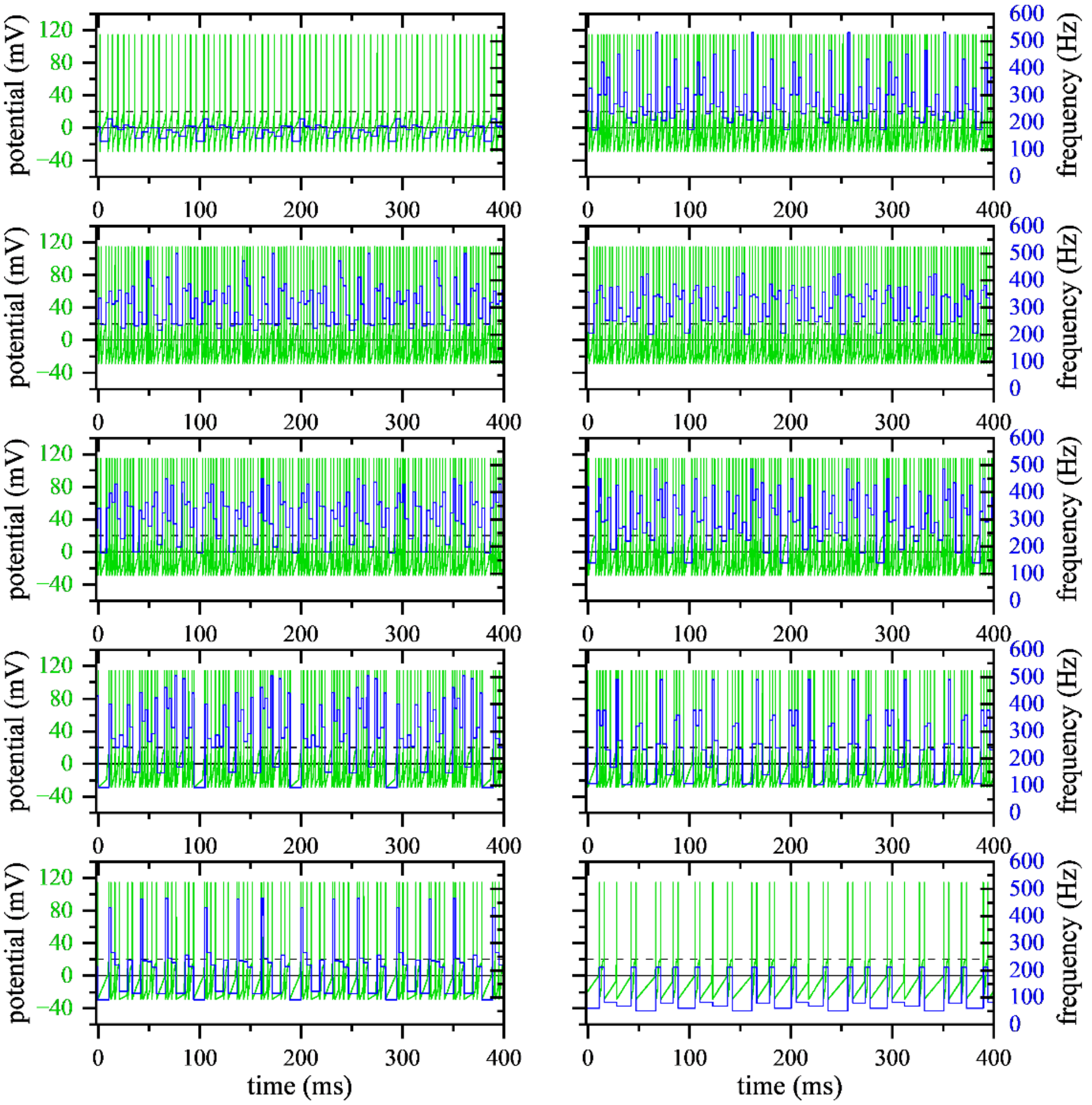

**Fig. S11.** Potential activity of the full positively feedback connected NS of 10 neurons. The initial condition is that the 3rd neuron triggers an action potential at 0.00 ms.

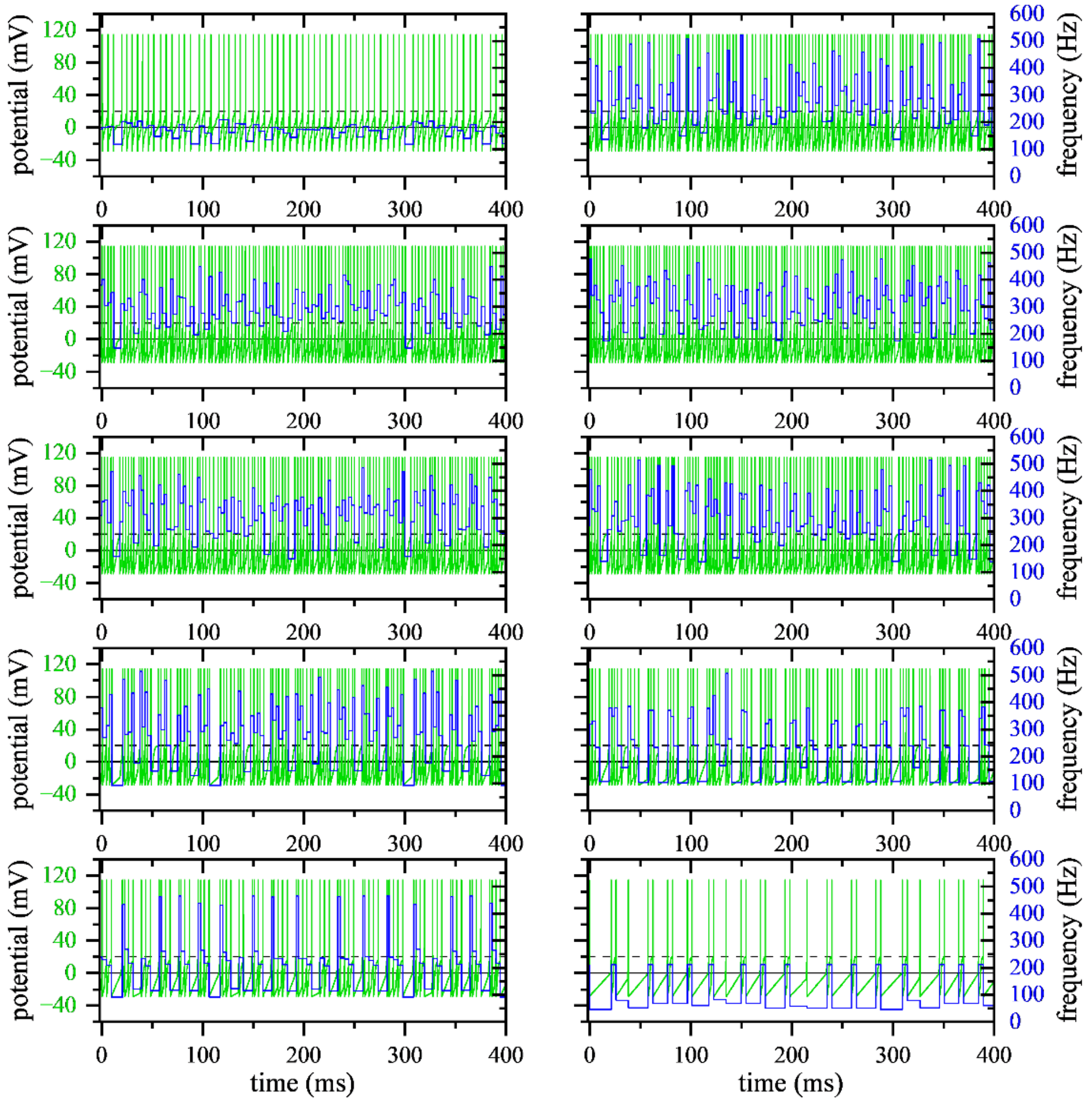

**Fig. S12.** Potential activity of the full positively feedback connected NS of 10 neurons. The initial condition is that the 4th neuron triggers an action potential at 0.00 ms.

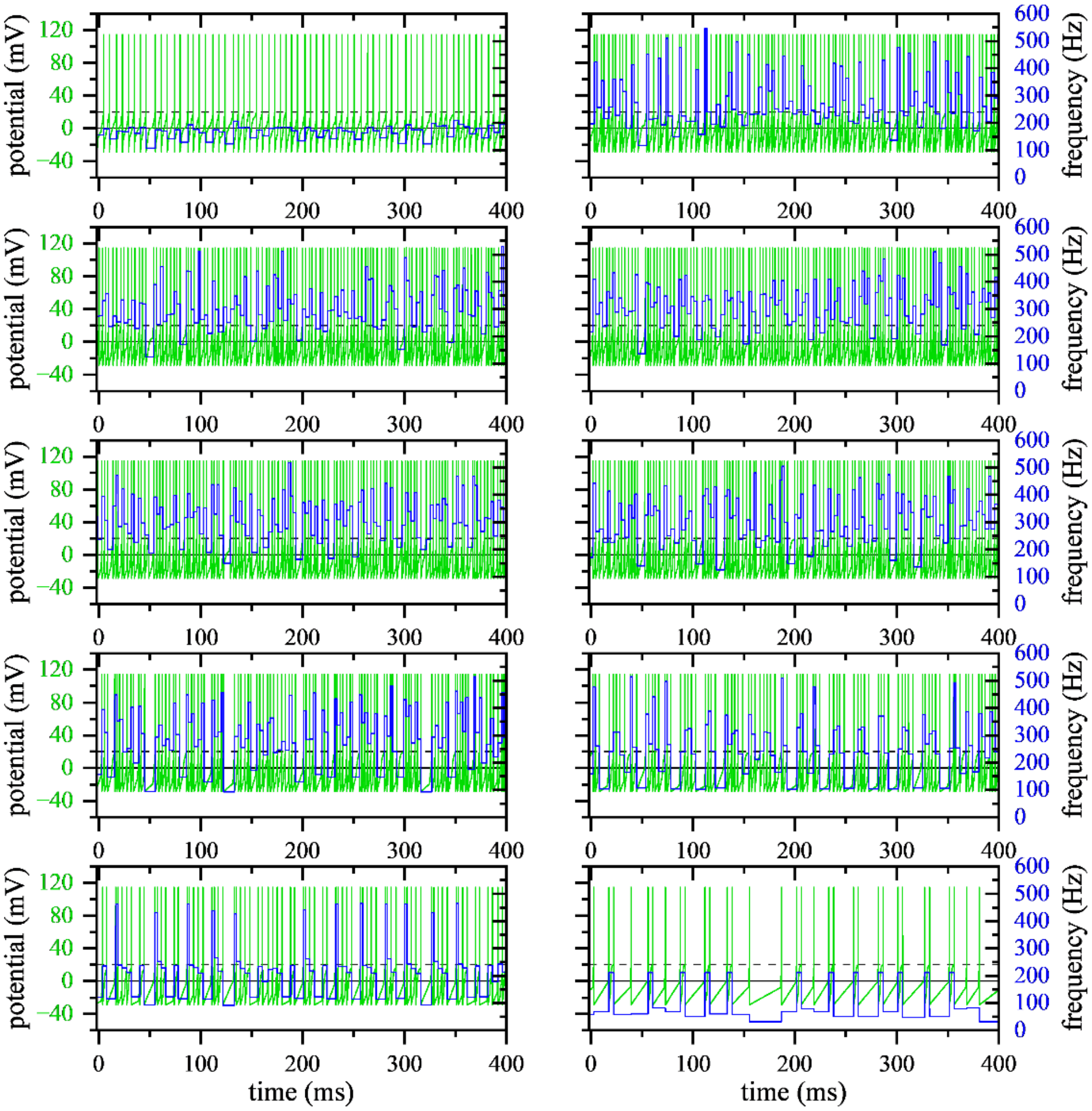

**Fig. S13.** Potential activity of the full positively feedback connected NS of 10 neurons. The initial condition is that the 5th neuron triggers an action potential at 0.00 ms.

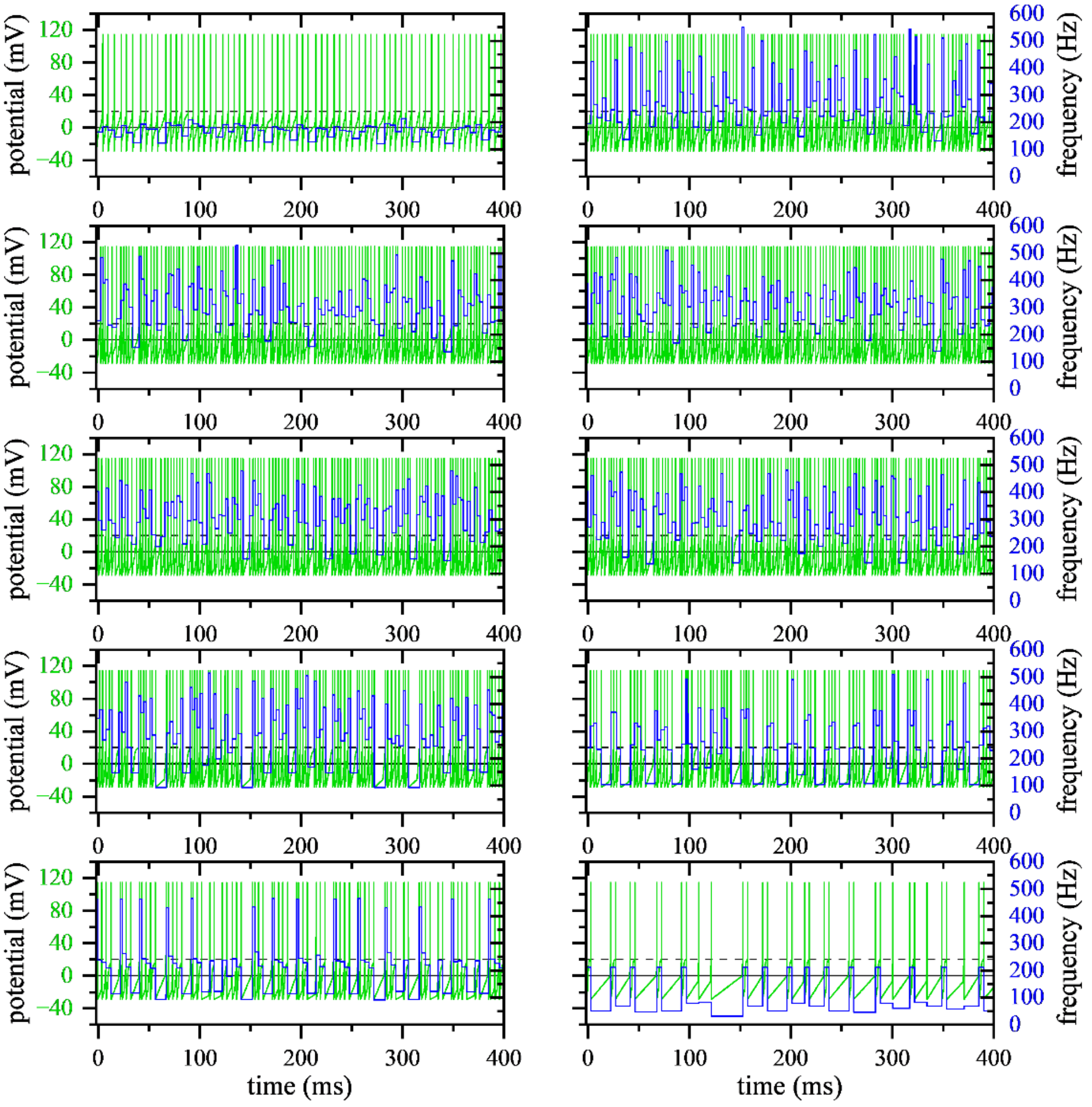

**Fig. S14.** Potential activity of the full positively feedback connected NS of 10 neurons. The initial condition is that the 6th neuron triggers an action potential at 0.00 ms.

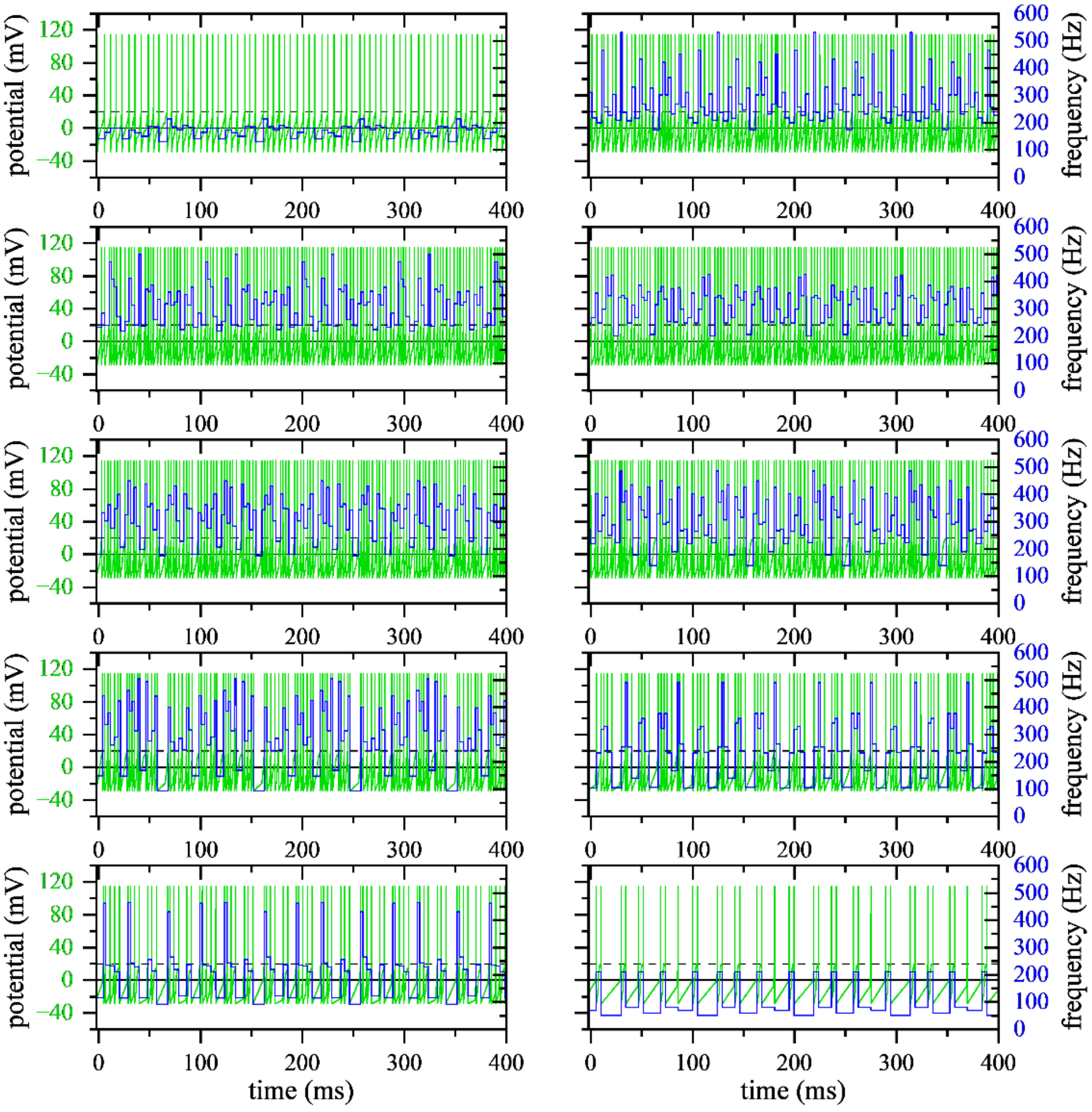

**Fig. S15.** Potential activity of the full positively feedback connected NS of 10 neurons. The initial condition is that the 7th neuron triggers an action potential at 0.00 ms.

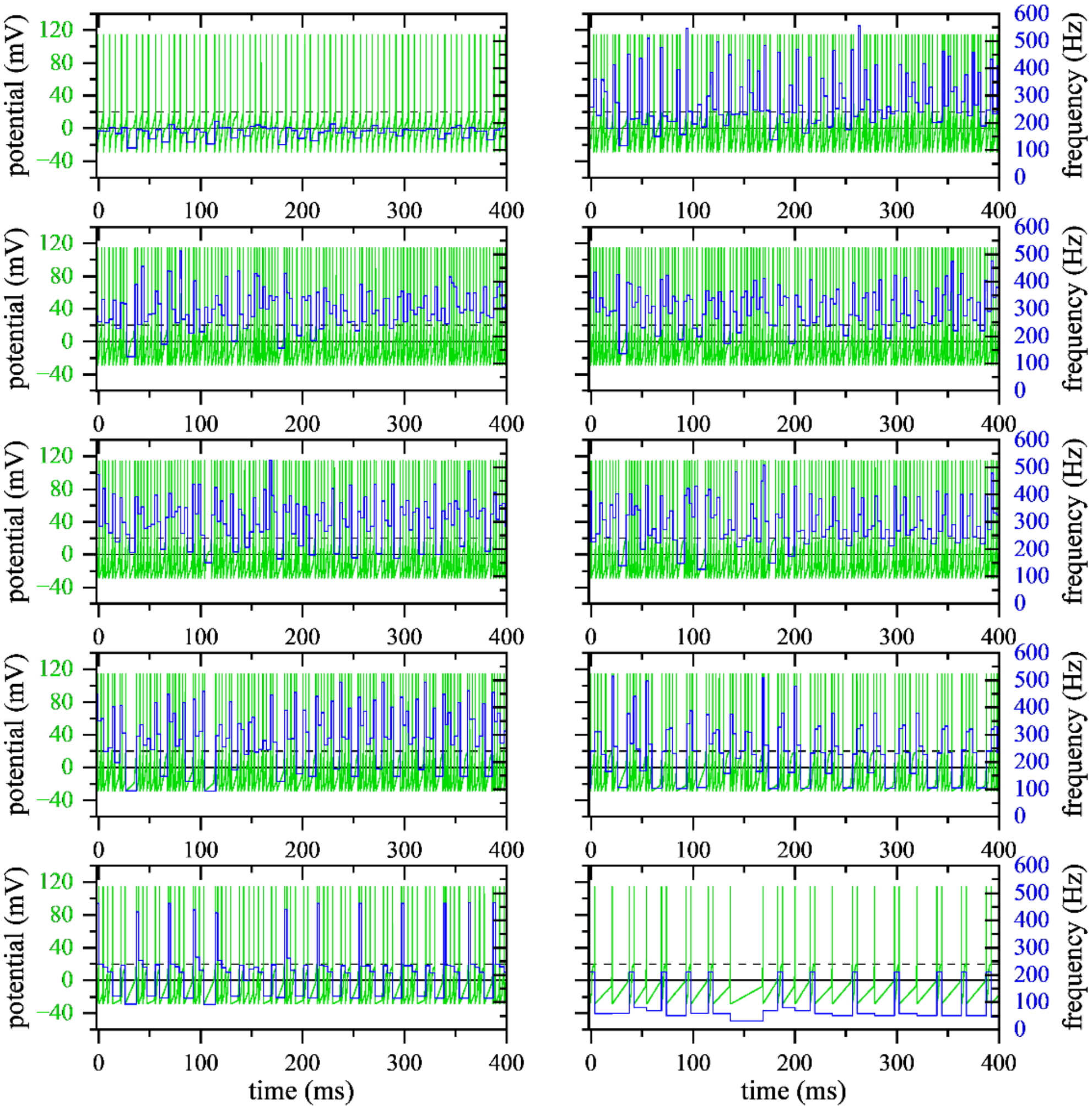

**Fig. S16.** Potential activity of the full positively feedback connected NS of 10 neurons. The initial condition is that the 8th neuron triggers an action potential at 0.00 ms.

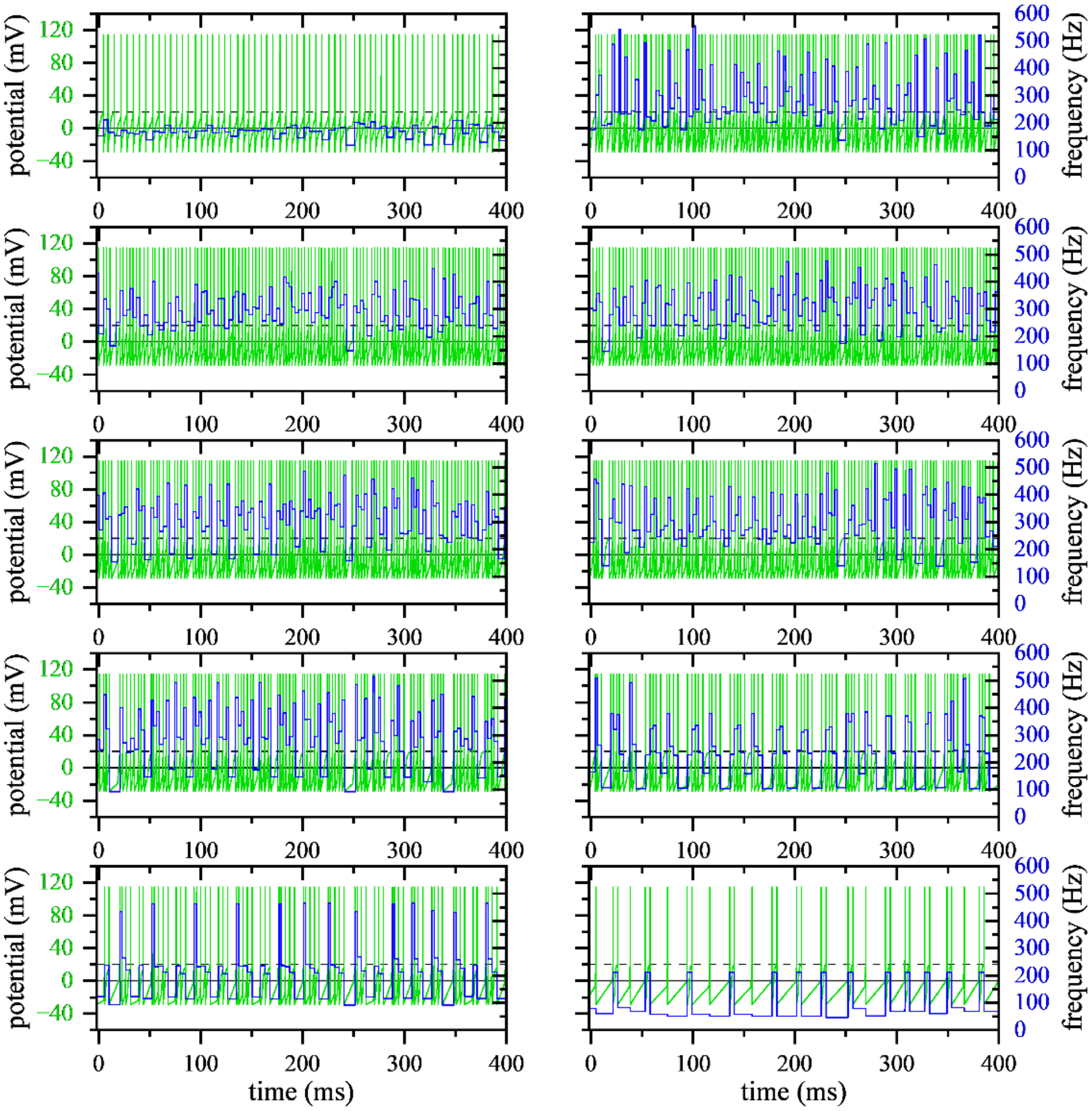

**Fig. S17.** Potential activity of the full positively feedback connected NS of 10 neurons. The initial condition is that the 9th neuron triggers an action potential at 0.00 ms.

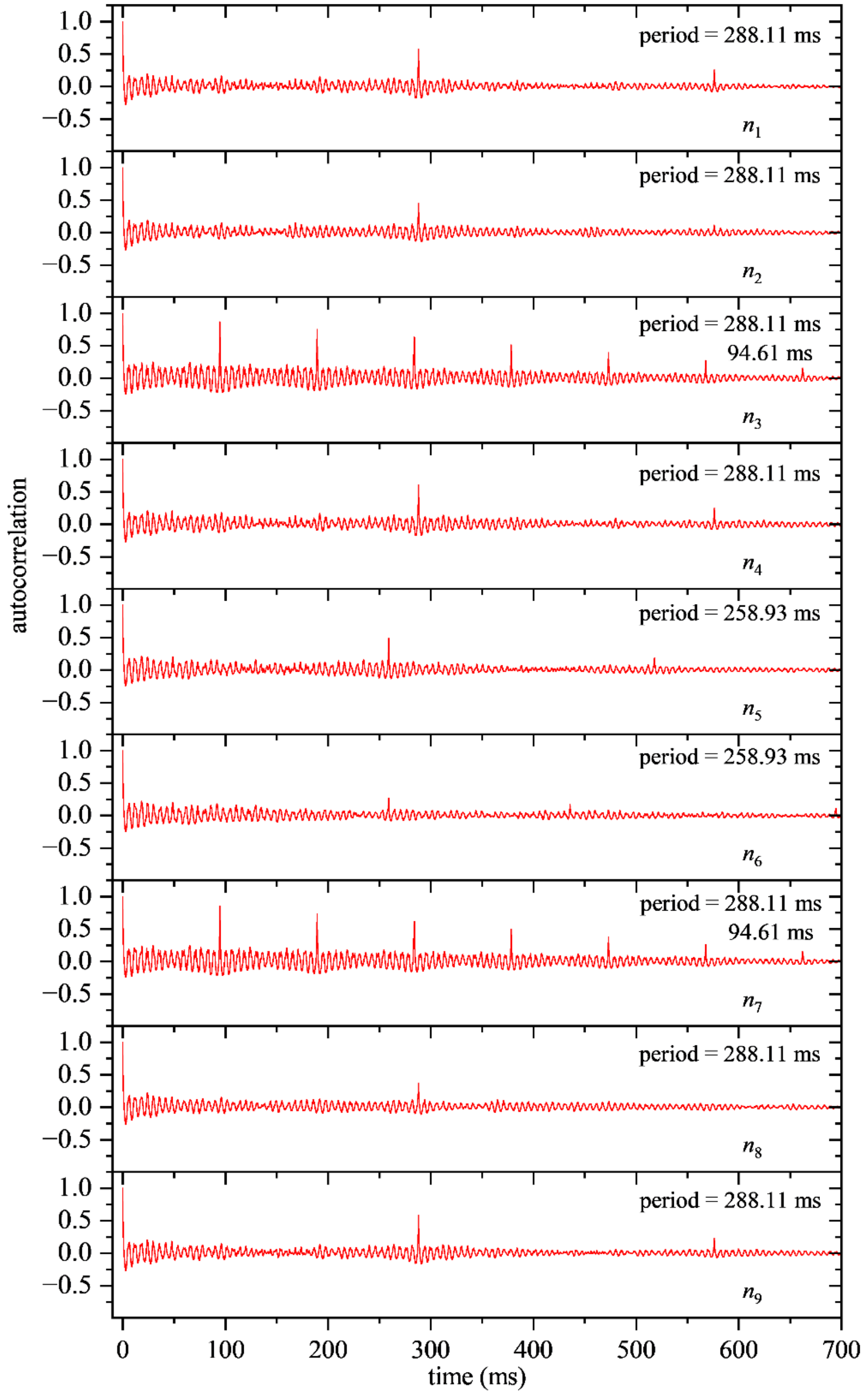

**Fig. S18.** The autocorrelation coefficient of the periodic potential waveform of full positively feedback connected NS of 10 neurons for different initial conditions. The initial condition $n_i$ is that the neuron $i$ triggers an action potential at 0.00 ms.

Electrophysiological simulations of full initial conditions for NS with five neurons

To characterize the information features of NS, the electrophysiological activities of full positively feedback NS with $n = 5$ and $q = 1$ in full initial conditions are simulated. The minimum difference between the elements of the initial condition matrix is 0.1 ms. Their upper limits are all less than 5 ms. For each initial condition, membrane potential activity of NS was simulated for up to 50 ms at the time resolution of 0.01 ms. A total of 221,551 groups of such simulations were performed. The connection matrix of the full positively feedback NS with 5 neurons is shown in Table S3. The original Data of above simulation are stored in Data S1.

**Data S1. (original data.rar)** The “original data.rar” file contains two folders, “input” and “output”. In the file of “input”, "1.csv" to "5.csv" are the information of five neurons in turn. "matrices.csv" contains information about all the initial condition matrices. "synapse_connection_matrix_1.csv" and "synapse_connection_matrix_2.csv" contains information about the synaptic connections between neurons. In the file of “output”, "membrane_potential_activity_1.csv" to "membrane_potential_activity_221551.csv" are the simulated information of 221551 groups of membrane potential activity in turn. Please refer to "readme.txt" for details.

| Index of connection | Index of neuron (axon) | Index of synapse (axon) | Index of neuron (dendrite) | Index of synapse (dendrite) | stimulus intensity coefficient $k_v$ |
|---|---|---|---|---|---|
| 1 | 1 | 5 | 2 | 1 | 400 |
| 2 | 2 | 5 | 3 | 1 | 400 |
| 3 | 2 | 6 | 1 | 2 | 80 |
| 4 | 3 | 4 | 4 | 1 | 400 |
| 5 | 3 | 5 | 2 | 2 | 160 |
| 6 | 3 | 6 | 1 | 3 | 80 |
| 7 | 4 | 3 | 5 | 1 | 400 |
| 8 | 4 | 4 | 3 | 2 | 240 |
| 9 | 4 | 5 | 2 | 3 | 160 |
| 10 | 4 | 6 | 1 | 4 | 80 |
| 11 | 5 | 3 | 4 | 2 | 320 |
| 12 | 5 | 4 | 3 | 3 | 240 |
| 13 | 5 | 5 | 2 | 4 | 160 |

**Table S3.** The connection matrix of the full positively feedback NS with 5 neurons.

Simplified NS dynamical system with different control parameters

By changing the control parameters $a_i$', $K_{ij}$' and $\Phi_i$' of the simplified NS dynamical system, different Phase portraits with beautiful fractal patterns can be generated. Figures. S19-S23 show the phase portraits of the full positively feedback NS with $n = 5$ for some control parameters, where $a_i$' is 1.00 ~ 1.00+0.25$i$,

$$\boldsymbol{K}_{\mathrm{V}}' = \begin{bmatrix} 0 & 0.4+0.01i & 0.3+0.01i & 0.2+0.01i & 0.1+0.01i \\ 0.5+0.01i & 0 & 0.4+0.01i & 0.3+0.01i & 0.2+0.01i \\ 0 & 0.5+0.01i & 0 & 0.4+0.01i & 0.3+0.01i \\ 0 & 0 & 0.5+0.01i & 0 & 0.4+0.01i \\ 0 & 0 & 0 & 0.5+0.01i & 0 \end{bmatrix}, \quad \text{(S.58)}$$

and $\Phi_i$' is 0.02 ~ 0.2. When plotting the phase portrait of neuron $j$, the rest of the $z_i$ ($i \neq j$) are all 0. The detailed visualization of the evolution process of the simplified NS dynamical system is presented in Movie S2.

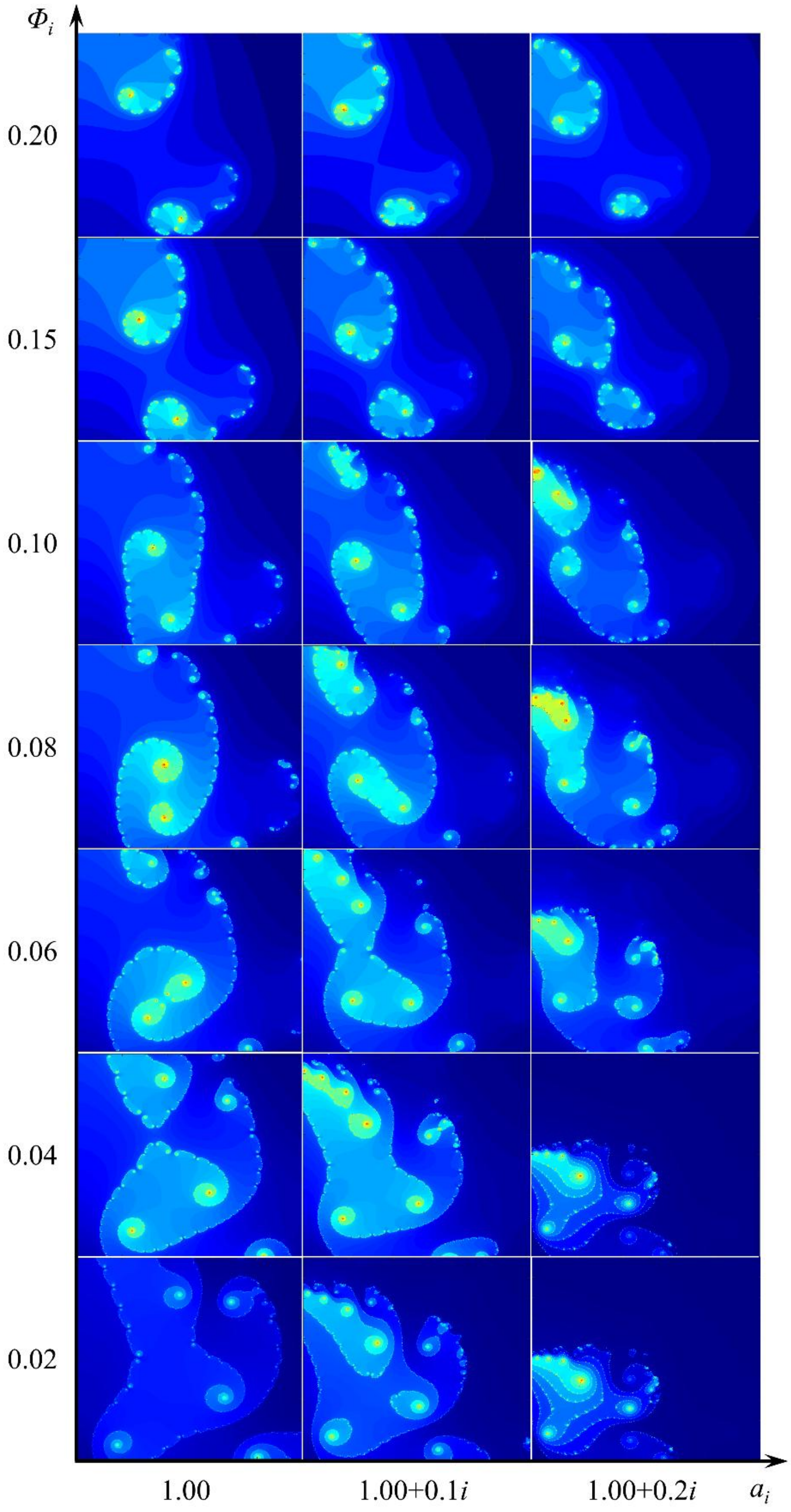

**Fig. S19.** Phase portraits of the simplified NS dynamical system with different control parameters for neuron 1. From blue to yellow and then to red, the number of iterations increased successively. The range of the coordinates of each phase portrait are all [0 5 ms].

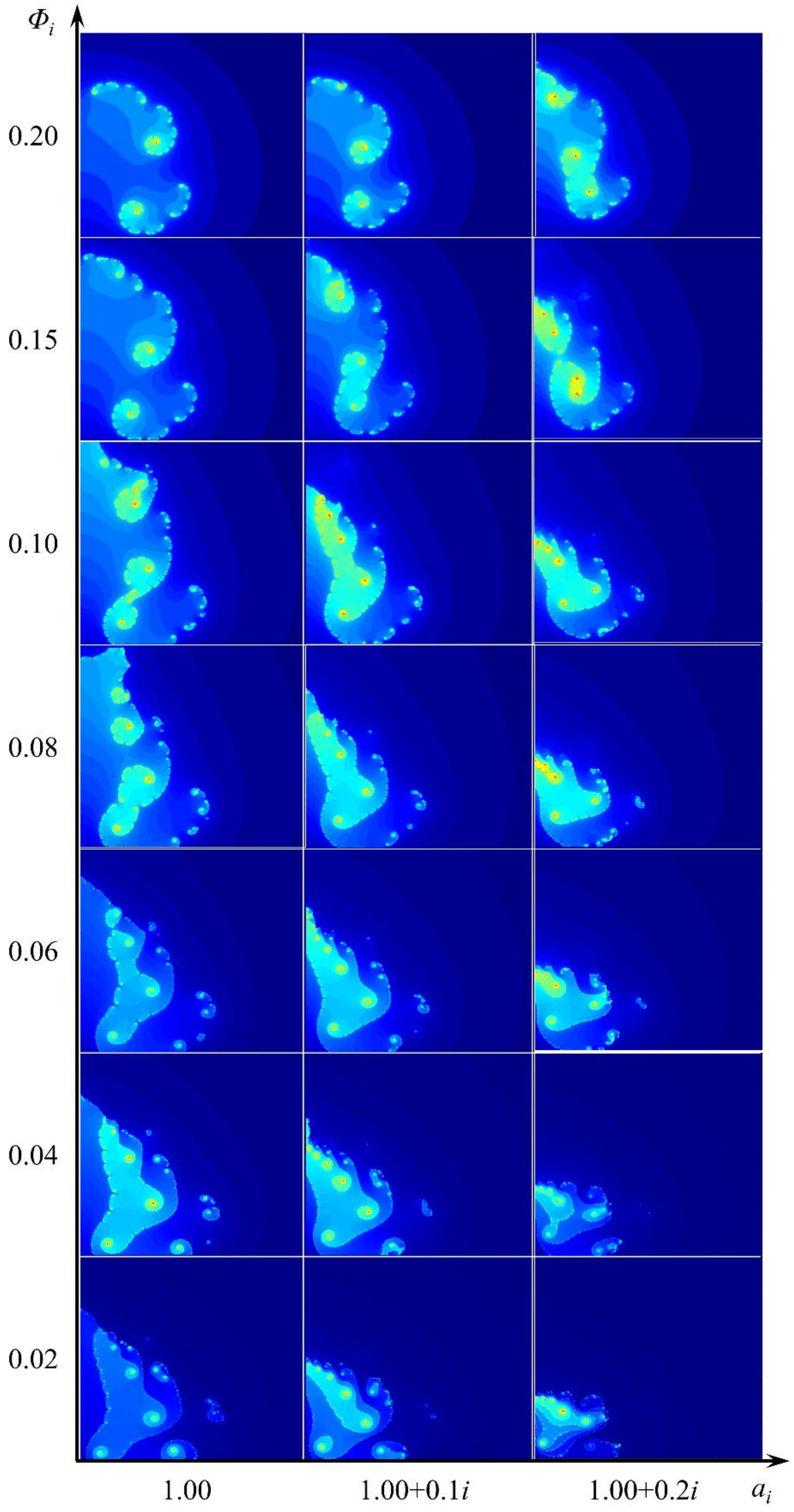

**Fig. S20.** Phase portraits of the simplified NS dynamical system with different control parameters for neuron 2. From blue to yellow and then to red, the number of iterations increased successively. The range of the coordinates of each phase portrait are all [0 5 ms].

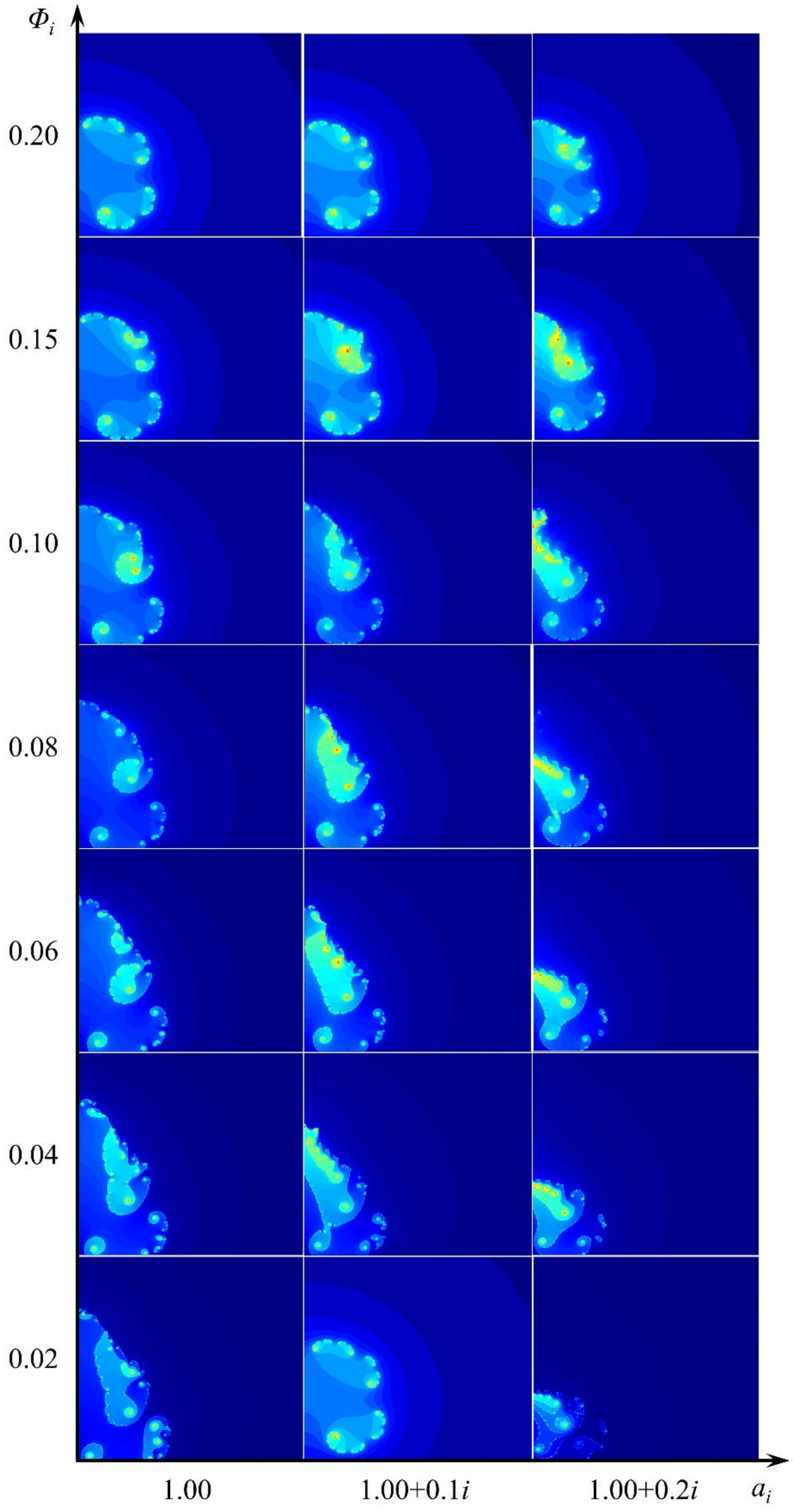

**Fig. S21.** Phase portraits of the simplified NS dynamical system with different control parameters for neuron 3. From blue to yellow and then to red, the number of iterations increased successively. The range of the coordinates of each phase portrait are all [0 5 ms].

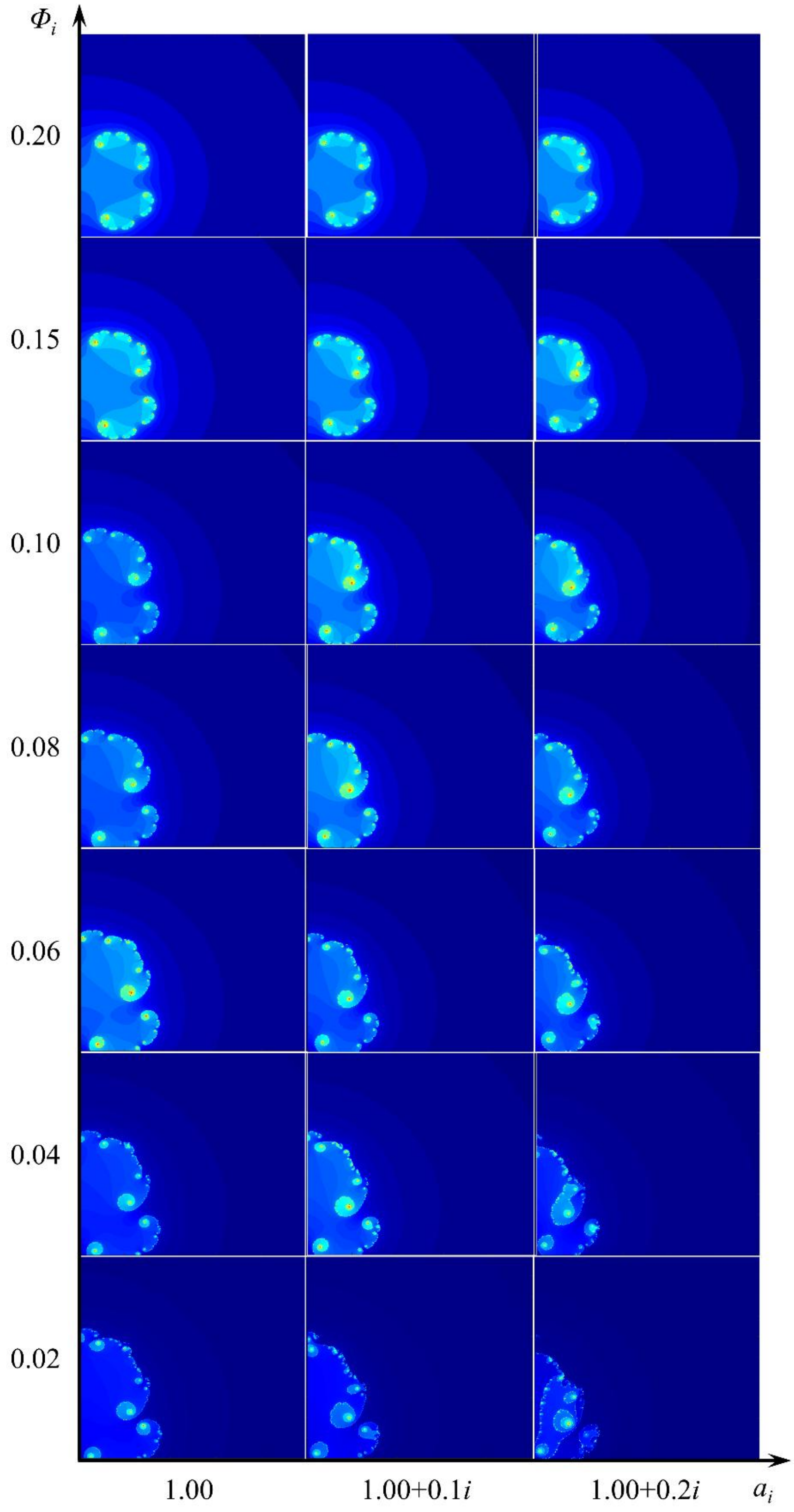

**Fig. S22.** Phase portraits of the simplified NS dynamical system with different control parameters for neuron 4. From blue to yellow and then to red, the number of iterations increased successively. The range of the coordinates of each phase portrait are all [0 5 ms].

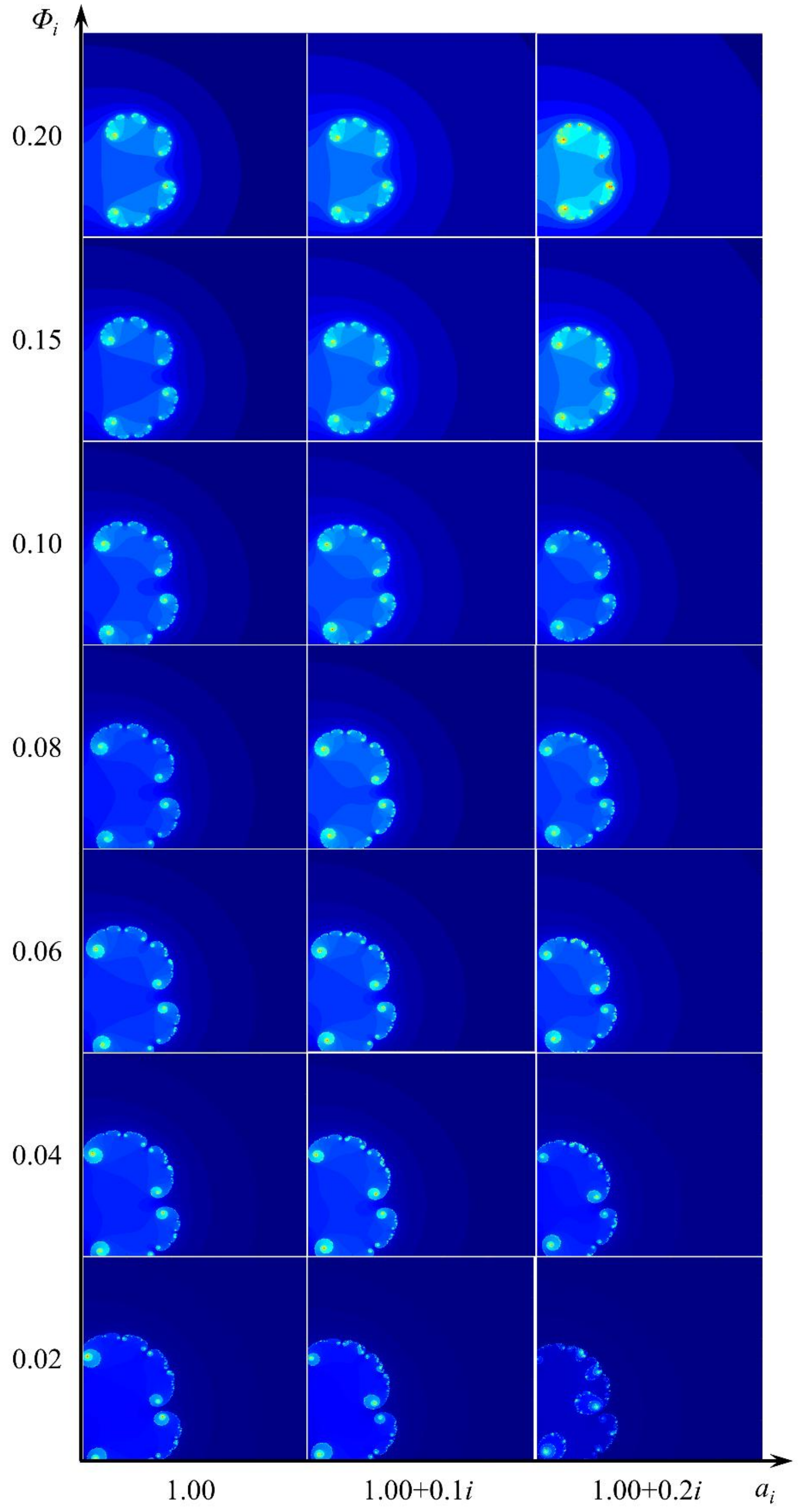

**Fig. S23.** Phase portraits of the simplified NS dynamical system with different control parameters for neuron 5. From blue to yellow and then to red, the number of iterations increased successively. The range of the coordinates of each phase portrait are all [0 5 ms].

Evaluation of the storage capacity for the HNCA

The storage capacity of the HNCA can be quantified to compare with other artificial devices. In addition to the widely used binary systems, the emerging two-dimensional material devices can generate many distinguishable states through regulation. The latest achievement of our team can regulate two-dimensional materials to generate up to thousands of states. Therefore, it is included in the comparison with HNCA here.

In addition to the various parameters of the NS, the storage capacity of the HNCA is affected by the temporal resolution $t_{min}$. In the ideal mathematical limit, the temporal resolution can be infinitesimal. In this case, HNCA can store an infinite amount of information. In the physical limit, the smallest temporal resolution is Planck time, $5.39\times10^{-44}$ s. These results are presented in Fig. 5a and 5b. Although the exact limits of biological temporal resolution remain unknown, the temporal resolution of neurons can be roughly estimated based on the timescale of ion transport in neuronal ion channels and may potentially reach values on the order of attoseconds to picoseconds. These results are presented in Fig. S24. The figure shows that the self-information of HNCA is also much higher than that of the artificial device by about 1 to 2 orders of magnitude when the temporal resolution is between attoseconds and picoseconds.

The storage capacity of the human brain can be simply estimated via HNCA. Assuming that the human brain with 86 billion neurons, each neuron has 5000 synapses. The time resolution is Planck time.

The first is the conservative estimate. In this case, the brain is simply sparsely linearly connected by a large number of small-scale ensembles. Therefore, the self-information of the brain is only a linear accumulation of the self-information of small-scale ensembles. Directly applying Equation (S.47), the self-information of the brain can be estimated to be about 8.24 PB.

The second is the aggressive estimation. In this case, the neural network in the brain is a complex fractal structure as shown in box IV of Fig. 4A. 86 billion neurons correspond to a four-layer fractal for a small-scale ensemble of 5 neurons, i.e., fractal of 5’’’×5’’×5’×5. After three recurrences through Equations (S.47) and (S.50), the self-information of the brain can be estimated to be about 7.48 EB (i.e., $7.48\times10^{6}$ TB).

Besides, the information compression ratio of HNCA can be simply estimated at a low temporal resolution. When $t_{min} = 1\times10^{-9}$, the self-information of the initial conditions for four neurons is $3.74\times10^{3}$ bit according to Equation (S.53) and the self-information of the NS of the four neurons is 198 bit. Therefore, the information compression ratio is $1 - (198 / 3.74\times10^{3}) = 94.7\%$. In smaller $t_{min}$ and NS with more neurons, the information compression ratio will be closer to 100%. It indicates that the brain has a very powerful ability of information compression and feature extraction.

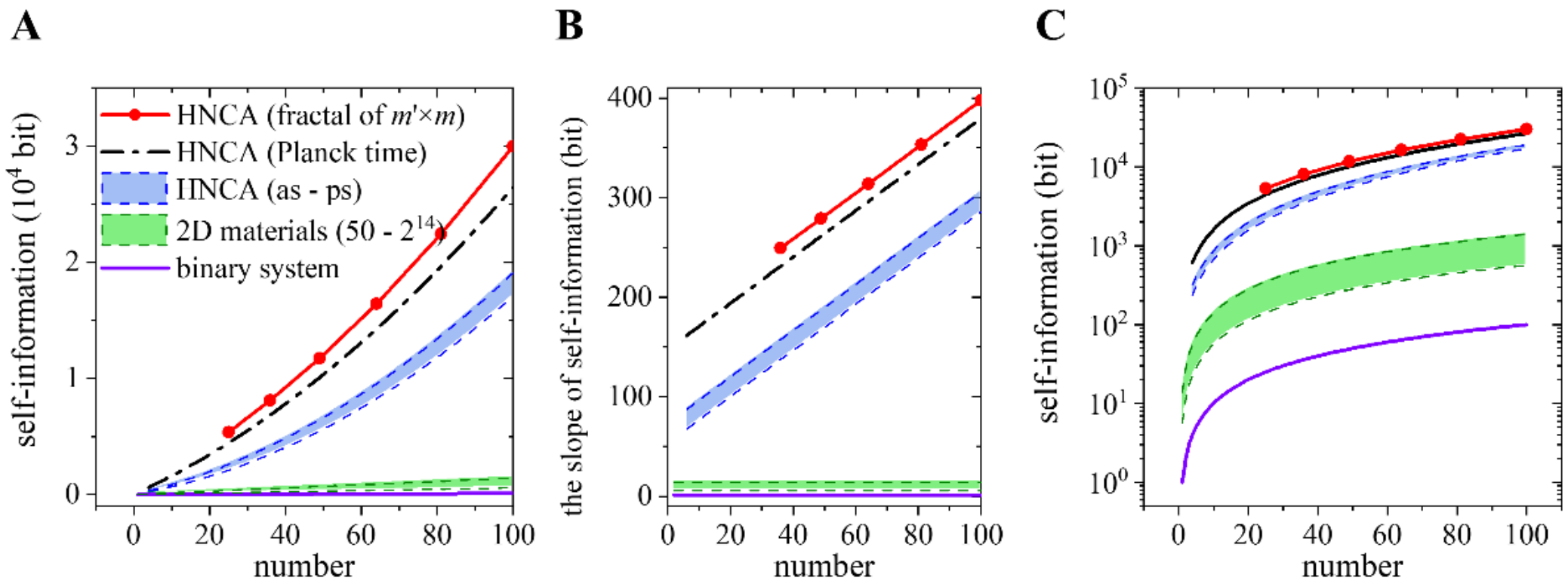

**Fig. S24.** Evaluation of storage capacity for HNCA. (**A**) Curves of the self-information of different systems as a function of the number of neurons or devices. The red line represents six memory NSs with the fractal of $m' \times m$ ($m$ = 5, 6, …, 10). The black line represents the memory NS with a Planck time as its temporal resolution. The blue area represents the temporal resolution of the memory NS from one attosecond to one picosecond. The green area represents a system of some neuromorphic two-dimension material devices, with a single device having 50 to $2^{14}$ distinguishable states(*39, 40, 42*). The purple line represents the binary system. (**B**) The slopes of the self-information curves in **A**. (**C**) The logarithmic coordinate of the self-information curves in **A**.

Evaluation of the computational power for the HNCA

The computational power of silicon-based devices is generally measured by floating point operations per second (FLOPS)(*33*). For CPU, GPU and other high performance chips, there are a set of well-established methods to test their computational power. According to the Table S4, the data of the latest generation high performance CPU and GPU and the top 10 supercomputer from various institutions or manufacturers are presented in Fig. 5C.

The computational power for floating-point operation of HNCA can be estimated by two equivalent methods according to Evaluation of the computational power for HNCA in **Materials and Methods**. In NSs with random action potential activity, Equation (S.54) is used for estimation. For the entire human brain, $FLOP_{HNCA} \in [10^3, 10^4]$ and $f_{max}$ = 1/ absolutely refractory period $\in$ [1000, 2000]. Therefore, the computational power of the human brain modeled by HNCA is estimated to range between 86 PFLOPS and 1.72 EFLOPS. Moreover, in this equivalent procedure, the precision of one floating-point operation is $\log_2(\tau_m / t_{min}) \in$ [34 bit, 138 bit]. In NSs with random action potential activity, Equation (S.55) is used for estimation. For the entire human brain, $I$ = 7.5 EB, $\mu$ = 64 bit and $T$ = 50 ms. Therefore, the computational power of the human brain modeled by HNCA is estimated to be 6.24 EFLOPS.

| Type | Model number | Computational power | Design power | Data sources |
|---|---|---|---|---|
| CPU | Intel® Xeon® 6745P Processor | 4.40 TFLOPS | 300 W | https://www.intel.com/content/www/us/en/ark.html#@Processors |
| CPU | AMD EPYC™ 9965 | 6.91 TFLOPS | 500 W | https://www.amd.com/en/products/processors/server/epyc/9005-series.html |
| GPU | AMD Radeon™ PRO W7900 Dual Slot | 61.3 TFLOPS | 295 W | https://www.amd.com/en/products/graphics/workstations.html |
| GPU | GeForce RTX 4090 | 82.5 TFLOPS | 450 W | https://www.nvidia.com/en-us/geforce/graphics-cards/40-series/rtx-4090/ |
| Super-computer | El Capitan | 2,746.38 PFLOPS | 29,581 kW | https://www.top500.org/lists/top500/2024/11/ |
| Super-computer | Frontier | 2,055.72 PFLOPS | 24,607 kW | https://www.top500.org/lists/top500/2024/11/ |
| Super-computer | Aurora | 1,980.01 PFLOPS | 38,698 kW | https://www.top500.org/lists/top500/2024/11/ |
| Super-computer | Eagle | 846.84 PFLOPS | / | https://www.top500.org/lists/top500/2024/11/ |
| Super-computer | HPC6 | 606.97 PFLOPS | 8,461 kW | https://www.top500.org/lists/top500/2024/11/ |
| Super-computer | Supercomputer Fugaku | 537.21 PFLOPS | 29,899 kW | https://www.top500.org/lists/top500/2024/11/ |
| Super-computer | Alps | 574.84 PFLOPS | 7,124 kW | https://www.top500.org/lists/top500/2024/11/ |
| Super-computer | LUMI | 531.51 PFLOPS | 7,107 kW | https://www.top500.org/lists/top500/2024/11/ |
| Super-computer | Leonardo | 306.31 PFLOPS | 7,494 kW | https://www.top500.org/lists/top500/2024/11/ |
| Super-computer | Tuolumne | 288.88 PFLOPS | 3,387 kW | https://www.top500.org/lists/top500/2024/11/ |

**Table S4.** The computational power and design power data of the latest generation high performance CPU and GPU and the top 10 supercomputer from various institutions or manufacturers.

Evaluation of the storage capacity and computational power for other organisms by HNCA

Due to the universality of HNCA, the storage capacity and computational power of the brain of other organisms can be estimated as shown in Fig. S25. The brain of the same organism has the same order of magnitude of storage capacity and computational power. In the brains of Caenorhabditis elegans(*38*), drosophila melanogaster(*19*), ant(*41*), mouse(*37*) and human, the human brain has an absolute lead, beating the mouse by 4 orders of magnitude. This benefits from the fact that the human brain has a large number of neurons and an ordered connection structure.

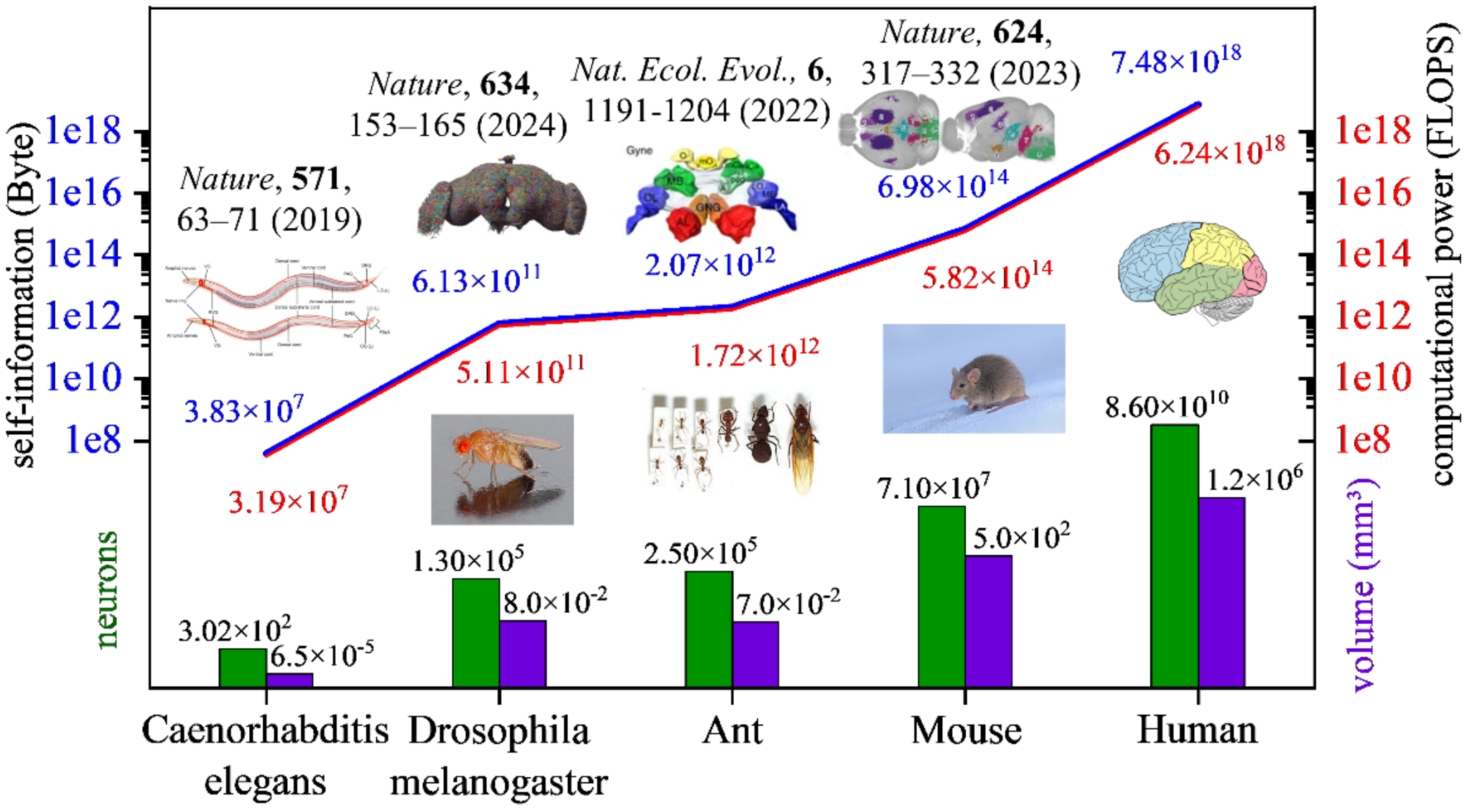

**Fig. S25.** Evaluation of the storage capacity and computational power for Caenorhabditis elegans(*38*), drosophila melanogaster(*19*), ant(*41*), mouse(*37*) and human by HNCA. The storage capacity is measured by self-information. The computational power is measured by double precision FLOPS. In the bar graph, green represents the number of neurons and purple represents the volume of the brain.

## References

1. D. Silver *et al.*, Mastering the game of Go with deep neural networks and tree search. *Nature* **529**, 484-489 (2016).
2. H. C, G. P, A. D, Updated Energy Budgets for Neural Computation in the Neocortex and Cerebellum. *Journal of Cerebral Blood Flow & Metabolism* **32**, 1222-1232 (2012).
3. K. Roy, A. Jaiswal, P. Panda, Towards spike-based machine intelligence with neuromorphic computing. *Nature* **575**, 607-617 (2019).
4. D. Kudithipudi *et al.*, Neuromorphic computing at scale. *Nature* **637**, 801-812 (2025).
5. S. Cajal, Estructura de los centros nerviosos de las aves. *Rev. Trim. Histol. Norm. Patol.* **1**, 1-10 (1888).
6. A. L. Hodgkin, A. F. Huxley, B. Katz, Measurement of current-voltage relations in the membrane of the giant axon of loligo. *The Journal of Physiology* **116**, 449-472 (1952).
7. A. L. Hodgkin, A. F. Huxley, Currents carried by sodium and potassium ions through the membrane of the giant axon of loligo. *The Journal of Physiology* **116**, 449-472 (1952).
8. A. L. Hodgkin, A. F. Huxley, The components of membrane conductance in the giant axon of loligo. *The Journal of Physiology* **116**, 473-496 (1952).
9. A. L. Hodgkin, A. F. Huxley, The dual effect of membrane potential on sodium conductance in the giant axon of loligo. *The Journal of Physiology* **116**, 497-506 (1952).
10. A. L. Hodgkin, A. F. Huxley, A quantitative description of membrane current and its application to conduction and excitation in nerve. *The Journal of Physiology* **116**, 500-544 (1952).
11. R. D. Traub, R. K. Wong, R. Miles, H. Michelson, A model of a CA3 hippocampal pyramidal neuron incorporating voltage-clamp data on intrinsic conductances. *Journal of neurophysiology* **66**, 635-650 (1991).
12. X. J. Wang, G. Buzsáki, Gamma oscillation by synaptic inhibition in a hippocampal interneuronal network model. *Journal of neuroscience* **16**, 6402-6413 (1996).
13. P. J. Basser, J. Mattiello, D. LeBihan, MRI of neuronal network structure, function, and plasticity. *Biophysical Journal* **66**, 259-267 (1994).
14. N. A. Steinmetz *et al.*, Neuropixels 2.0: A miniaturized high-density probe for stable, long-term brain recordings. *Science* **372**, eabf4588 (2021).
15. N. J. Sofroniew, D. Flickinger, J. King, K. Svoboda, A large field of view two-photon mesoscope with subcellular resolution for in vivo imaging. *eLife* **5**, e14472 (2016).
16. C. Stringer, M. Pachitariu, Analysis methods for large-scale neuronal recordings. *Science* **386**, eadp7429 (2024).
17. G. A. Ascoli, D. E. Donohue, M. Halavi, NeuroMorpho.Org: A Central Resource for Neuronal Morphologies. *Journal of Neuroscience* **27**, 9247-9251 (2007).
18. J. G. White, E. Southgate, J. N. Thomson, S. Brenner, The structure of the nervous system of the nematode Caenorhabditis elegans. *Philos Trans R Soc Lond B Biol Sci* **314**, 1-340 (1986).
19. A. Lin *et al.*, Network statistics of the whole-brain connectome of Drosophila. *Nature* **634**, 153-165 (2024).
20. N. L. Turner *et al.*, Reconstruction of neocortex: Organelles, compartments, cells, circuits, and activity. *Cell* **185**, 1082-1100 (2022).
21. C. R. Gamlin *et al.*, Connectomics of predicted Sst transcriptomic types in mouse visual cortex. *Nature* **640**, 497-505 (2025).
22. O. Sporns, G. Tononi, R. Kötter, The human connectome: a structural description of the human brain. *PLoS Comput. Biol.* **1**, e42 (2005).

23. M. Naddaf, Europe spent €600 million to recreate the human brain in a computer. How did it go? *Nature* **620**, 718-720 (2023).
24. O. Sporns, R. Kötter, Motifs in brain networks. *PLOS Biol.* **2**, e369 (2004).
25. L. Luo, Architectures of neuronal circuits. *Science* **373**, eabg7285 (2021).
26. T. M. B. Jr *et al.*, Nanoconnectomic upper bound on the variability of synaptic plasticity. *eLife* **4**, e10778 (2015).
27. W. Rall, Branching dendritic trees and motoneuron membrane resistivity. *Experimental neurology* **1**, 491-527 (1959).
28. Gonzalez, Alvaro, Measurement of Areas on a Sphere Using Fibonacci and Latitude-Longitude Lattices. *Mathematical Geosciences* **42**, 49-64 (2010).
29. K. I. Park, M. Park, in *Fundamentals of probability and stochastic processes with applications to communications*. (Springer, Cham, 2018), pp. 165-168.
30. G. Buzsáki, A. Draguhn, Neuronal Oscillations in Cortical Networks. *Science* **304**, 1926-1929 (2004).
31. R. J. Kosinski, A literature review on reaction time. *Clemson University* **10**, 337-344 (2008).
32. C. E. Shannon, A mathematical theory of communication. *The Bell System Technical Journal* **27**, 379-423 (1948).
33. K. D. J, *Parameters of a Stochastic Process*. Computer system capacity fundamentals (National Bureau of Standards, US Department of Commerce, 1974).
34. A. Bérut *et al.*, Experimental verification of Landauer’s principle linking information and thermodynamics. *Nature* **483**, 187-189 (2012).
35. Y. Jun, M. Gavrilov, J. Bechhoefer, High-Precision Test of Landauer’s Principle in a Feedback Trap. *Phys. Rev. Lett.* **113**, 190601 (2014).
36. P. Lennie, The Cost of Cortical Computation. *Current biology* **13**, 493-497 (2003).
37. Z. Yao *et al.*, A high-resolution transcriptomic and spatial atlas of cell types in the whole mouse brain. *Nature* **624**, 317-332 (2023).
38. S. J. Cook *et al.*, Whole-animal connectomes of both Caenorhabditis elegans sexes. *Nature* **571**, 63-71 (2019).
39. Z. Zheng *et al.*, Unconventional ferroelectricity in moiré heterostructures. *Nature* **588**, 71-76 (2020).
40. X. W. K. Yasuda, K. Watanabe, T. Taniguchi, P. Jarillo-Herrero, Stacking-engineered ferroelectricity in bilayer boron nitride. *Science* **372**, 1458-1462 (2021).
41. Q. Li *et al.*, A single-cell transcriptomic atlas tracking the neural basis of division of labour in an ant superorganism. *Nature ecology & evolution* **6**, 1191-1204 (2022).
42. D. Sharma *et al.*, Linear symmetric self-selecting 14-bit kinetic molecular memristors. *Nature* **633**, 560-566 (2024).